\documentclass[fleqn,usenatbib]{mnras}

\usepackage[T1]{fontenc}

\usepackage{graphicx}   

\usepackage{amsmath}
\usepackage{amssymb}
\usepackage{mathptmx}
\usepackage{txfonts}

\usepackage{float}
\usepackage{color}
\usepackage{enumitem}

\title[Formation of asymmetric arms in barred galaxies]{Formation of asymmetric arms in barred galaxies}

\author[P. S\'anchez-Mart\'{i}n et al.]{
P. S\'anchez-Mart\'{i}n,$^{1}$\thanks{E-mail: patricias@unizar.es}
C. Garc\'{\i}a-G\'omez,$^{2}$
J. J. Masdemont$^{3}$
and M. Romero-G\'{o}mez$^{4}$
\\
$^{1}$ IUMA, Universidad de Zaragoza, Dept. de Matem\'atica Aplicada, Pedro Cerbuna, 12, 50009 Zaragoza, Spain\\
$^{2}$D.E.I.M, Universitat Rovira i Virgili, Avd. Pa\"isos Catalans, 26, 43007 Tarragona, Spain\\
$^{3}$IEEC \&  Universitat Polit\`{e}cnica de Catalunya, Dept. de Matem\`{a}tiques, Diagonal 647, E08028 Barcelona, Spain\\
$^{4}$Institut de Ci\`{e}ncies del Cosmos (ICCUB), Universitat de Barcelona (IEEC-UB), Mart\'{i} i Franqu\`{e}s 1, E08028 Barcelona, Spain 
}

\date{Accepted XXX. Received YYY; in original form ZZZ}

\pubyear{2023}

\begin{document}
\label{firstpage}
\pagerange{\pageref{firstpage}--\pageref{lastpage}}
\maketitle


\begin{abstract}

We establish a dynamical mechanism to explain the origin of the asymmetry between the arms observed in some barred disk galaxies, where one of the two arms emanating from the bar ends is very well defined, while the second one displays a ragged structure, extending between its ridge and the bar. To this purpose, we study the invariant manifolds associated to the Lyapunov periodic orbits around the unstable equilibrium points at the ends of the bar. Matter from the galaxy center is transported along these manifolds to the periphery, forming this way the spiral arms that emanate from the bar ends. If the mass distribution in the galaxy center is not homogeneous, because of an asymmetric bar with one side stronger than the other, or because of a non-centered bulge, the dynamics about the two unstable Lagrange points at the ends of the bar will not be symmetric as well. One of their invariant manifolds becomes more extended than the other, enclosing a smaller section and the escaping orbits on it are fewer and dispersed in a wider region. The result is a weaker arm, and more ragged than the one at the other end of the bar.

\end{abstract}






\section{Introduction}

\begin{figure}
  \centering
  \includegraphics[scale=0.37]{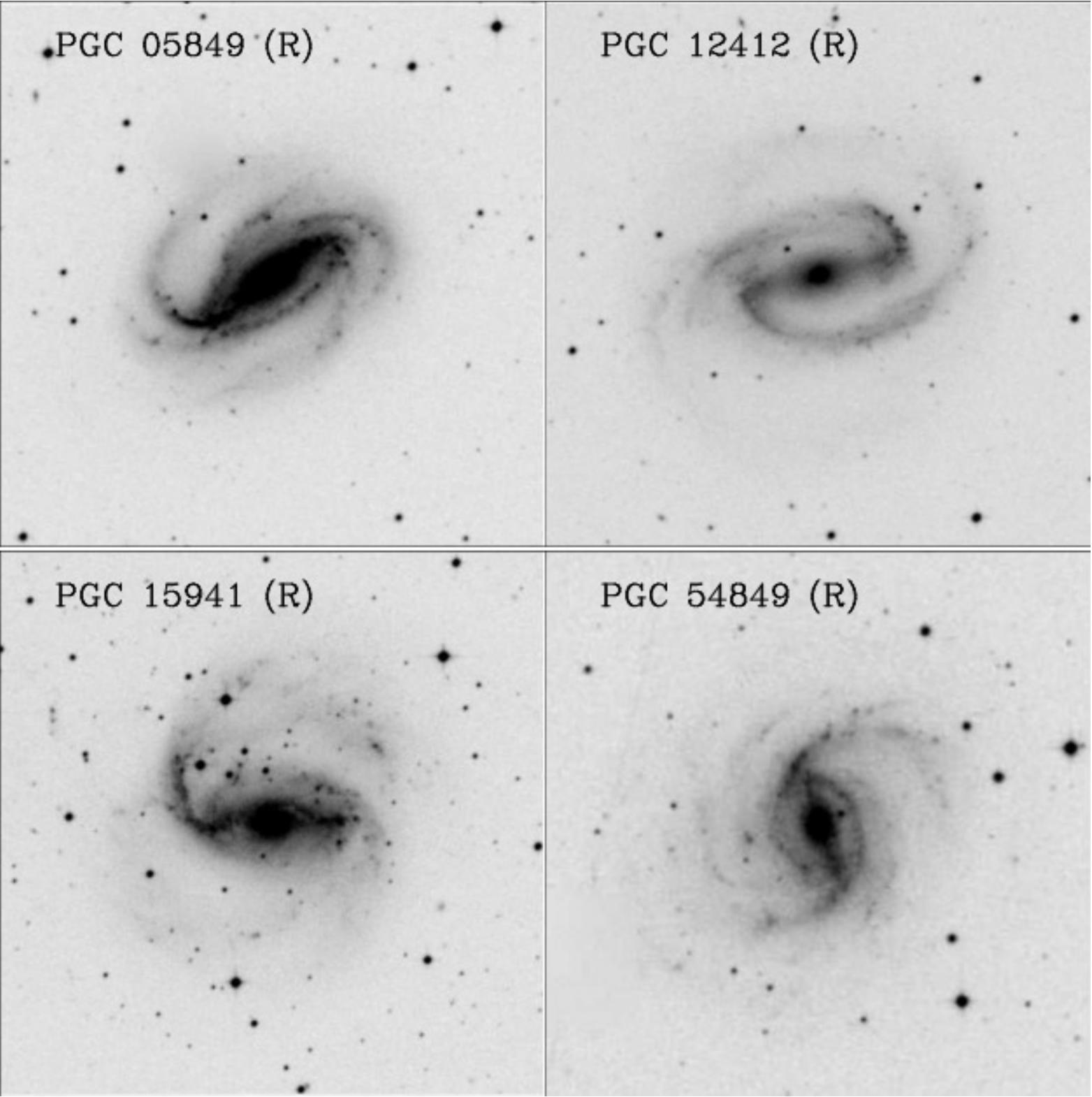}
  \centering
  \caption{Images of some strongly barred galaxies showing the asymmetric spiral arm patterns in the R-filter from the STScI Digitized Sky Survey. In the top row we show the galaxies PGC 05849 (NGC 0613), PGC 12412 (NGC 1300) and in the bottom one PGC 15941 (NGC 1672) and PGC 54849 (NGC 5921).}
  \label{fig:Barred}
\end{figure}

A high fraction of disk galaxies appear to be barred (74\% to 85\% in automatic classifications or 36\% to 63\% with alternative methods~\citep{Lee2019}).
These authors also suggest that strongly barred galaxies, classified as SBs, are preponderant in late-type galaxies.
Most of these strongly barred galaxies appear to be asymmetric under a simple visual inspection. They usually present a strong bar and two spiral arms emanating from the bar ends, and frequently some inner rings as well. However, a detailed visual inspection of these galaxies, available in the STScI Digitized Sky Survey or in the Sloan Digital Sky Survey (SDSS) \citep{Bundy2015}, reveals that some barred galaxies exhibit important asymmetries in their spiral structure. While one of the two spiral arms is long and strongly defined, the second arm shows a ragged structure, and the matter is distributed irregularly between the ridge of the spiral arm and the bar. A non-symmetric distribution of matter in the galaxy disc can be a natural outcome, since disc galaxies may be formed through a combination of secular evolution and violent events, including smooth accretion, disc instabilities and minor and major mergers \citep[e.g.][]{Tonini2016}. In Fig.~\ref{fig:Barred} we show some examples of galaxies showing asymmetric discs.

In this paper we study the dynamic response of an asymmetric mass distribution on the orbital structure of barred galaxies, modelling the asymmetry of the central parts by a slightly off-centered bulge which, viewing the galaxy as a whole, represents a simple but realistic model of an asymmetric bar. In this scenario, the Lagrange invariant points located at the bar ends will present also an asymmetric orbital structure, as initially shown in \citet{Colin1989}. 
The new step in this work is to additionally study the unstable periodic orbits around these Lagrange points and the associated unstable manifolds which provide escape routes for orbits which can transport material from the central regions to the outer parts of the discs  \citep{Romero1, Warps}. These unstable manifolds are the backbones of the spiral arms emanating from the bar ends. In the case of asymmetric Lagrange points, however, the unstable manifolds display also important differences. The orbits following one of these manifolds are close together and can explain the presence of a strong spiral arm. On the other hand, the second manifold has a more open orbital structure, with its escaping orbits dispersed in a wider zone, causing the ragged structure of the second arm. Thus, the orbital structure of the unstable Lagrange points is able to explain the asymmetric phenomenology present in some of the strongly barred galaxies.

The paper is organized as follows: in Section~\ref{sec:analysis} we quantify the level of asymmetry in asymmetric barred galaxies using a two-dimensional Fourier transform method. In Sections~\ref{sec:model} and \ref{sec:dynamics} we describe the asymmetric barred galaxy model used in this work and show the orbital analysis and invariant manifolds, respectively. Different approaches are summarized and conclusions are given in Section~\ref{sec:disc}.


\section{Analysis of the bar asymmetry}
\label{sec:analysis}

In this paper, we model the galactic asymmetric mass distribution using a classical galactic model with three symmetric components (disc, bar and bulge) but displacing slightly the bulge from the galaxy centre. This results in a total mass distribution biased towards one side of the bar. Many barred galaxies show some kind of asymmetries in their inner mass distribution with one side of the bar stronger than the other. An example is the galaxy NGC 1300, studied by \citet{Patsis2010}. They generated a numerical model of the potential using K images of the galaxy. The resultant model was clearly asymmetric, containing a bar with a stronger arm. This model was used to study the orbits associated with this asymmetric barred structure. Many barred galaxies show similar asymmetries. Some of these galaxies are shown in Fig.~\ref{fig:Barred}. In order to quantify these asymmetries, we can analyze in detail the mass distribution of one of the galaxies showing this phenomenology, namely the galaxy PGC 70419 (NGC 7479).
For this purpose, we use a galaxy image from the SDSS survey in the infrared z-filter. Assuming a constant $M/L$ ratio, the light distribution is a good tracer of the mass distribution in the disk. The image is previously cleaned from background stars and then deprojected using the FFT method \citep{GarciaGomez2004} giving a position angle (PA) of $38^\circ$ and an inclination angle of $45^\circ$. 

The image is then decomposed in its Fourier components using a technique first introduced by  \citet{Considere1982, Iye1982}, and further developed by \citet{GarciaGomez2017}. For deprojected image of the galaxy $I(u,\theta)$, where $u=\ln(r)$, we calculate the two-dimensional Fourier transform defined as
\begin{equation}
    A(p,m)=\int_{u_{min}}^{u_{max}} \int_{0}^{2\pi}I(u,\theta)e^{i(pu+m\theta)}\,d\theta du
\end{equation}
Where $m$ is the azimuthal frequency, associated with the multiplicity of the structures, i.e., the number of arms while $p$ is the radial frequency, associated to the pitch angle of the structure $i$, through the relation
\begin{equation}
    p=-\frac{m}{\tan(i)}
\end{equation}
In this way, the $m=1$ spectrum contains the spiral components with no symmetry, the $m=2$ spectrum the components with a periodicity of $\pi$ radians or bisymmetric signals, and so on for the rest of the $m$ frequencies. Each of the azimuthal components $m=1,2,...$ can be further decomposed in its radial components using a Gaussian fit to the modulus and keeping the phase constant as follows:
\begin{equation}
    \mid A(p,m) \mid = \sum_{j=1}^{N_{g}} C_j \exp{-\frac{(p-p_j)^2}{2\sigma_j^2}}.
\end{equation}
In this relation, $p_j$ represents the central frequency of the Gaussian, $\sigma_j$ its dispersion, and $C_j$ its amplitude. The number of Gaussians used in each fit, $N_g$, will depend on the complexity of the spectrum.

\begin{figure}
  \centering
    \includegraphics[scale=0.37]{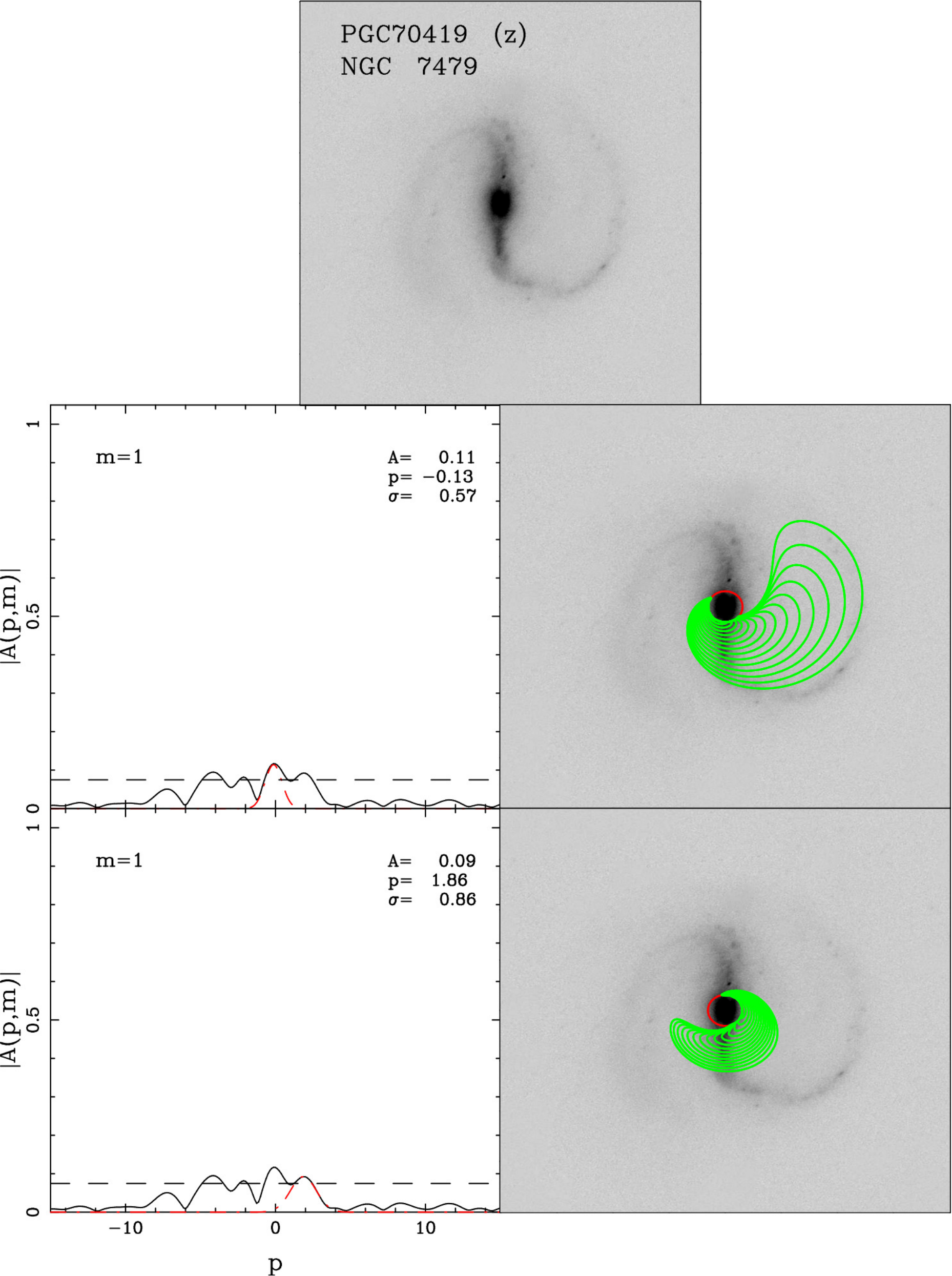}
  \caption{Analysis of the $m=1$ component of the SDSS galaxy image of NGC 7479 in the z-filter. The upper panel shows the deprojected image of the galaxy. On the left of the middle and lower panels we show the modulus of the $m=1$ component of the Fourier spectrum, and with red-dashed lines two Gaussian components fitted to this modulus. In the respective right panels, we show in green the density distribution associated to each of these single components superimposed on the galaxy image.}
  \label{fig:NGC7479}
\end{figure}

In the upper panel of Fig.~\ref{fig:NGC7479} we show the deprojected SDSS galaxy image of NGC~7479 using the z-filter. Note that one of its arms is very pronounced, while the opposite arm appears diffuse. On the left of the middle and lower panels of this figure we show the modulus of the Fourier spectrum of the $m=1$ component which is associated to the asymmetries in the light distribution. The modulus of the Fourier components are normalized to the modulus of the stronger component, which in this case is the $m=2$ containing the bisymmetric signals of the strong bar and the spiral arms. The $m=1$ spectrum shown here contains the spiral components responsible for  the asymmetries in the mass distribution. The relative low values of the $m=1$ in this scale shows that the mass asymmetry is a second order effect for this galaxy. In red we superpose two of the Gaussian components into which this signal is decomposed. These Gaussian components can be transformed back to obtain the density distribution associated to each particular spiral mode. The density distributions of these spiral components are presented in the respective right panels in green, superposed on the galaxy image. The isocontours come from the normalized density corresponding to the Gaussian components
and show that the asymmetries are related to the southern end of the bar and the strong arm emanating from its end. This shows that the galaxy has an asymmetric mass distribution, biased to the southern part of the galaxy.

\section{Characteristics of the Galactic Model}
\label{sec:model}

The equations of motion of the classical galactic model \citep[see e.g.][]{Pfenn, Skokos, Romero1, Warps} describe the movement of a particle in a gravitational potential $\phi$ . In the rotating frame, the equations of motion are described by
\begin{equation}\label{eqn:motion}
 \ddot{\mathbf{r}} = -2(\mathbf{\Omega}_\text{p}\times\dot{\mathbf{r}}) - \mathbf{\Omega}_\text{p}\times(\mathbf{\Omega}_\text{p}\times\mathbf{r}) -\nabla{\phi},
\end{equation}
where $\mathbf{r}=(x,\,y,\,z)$ is the position of the particle, $\phi$ is the total gravitational potential of the system and $\mathbf{\Omega}_\text{p}$ is the angular velocity of the bar around the $z$-axis, $\mathbf{\Omega}_\text{p} = (0,0,\Omega)$. The origin of the reference frame is located at the center of mass of the system and the frame is aligned with the main axis of the bar.

The potential model $\phi$ used in this paper is a combination of three analytical components:  an axisymmetric Miyamoto-Nagai disc with potential $\phi_d$ \citep{Miyamoto}, an ellipsoid Ferrers bar with potential $\phi_b$ \citep{Ferrers} and a bulge structure represented by a Plummer spheroid potential $\phi_{bl}$ \citep{Plummer1911}. The total potential is the addition of these three components, $\phi = \phi_d + \phi_b+\phi_{bl}$. 

The disc potential is described by the equation

\begin{equation}\label{eqn:Miyamoto}
 \phi_d = - \frac{GM_d}{\sqrt{R^2+(A+\sqrt{B^2+z^2})^2}},
\end{equation}
were $R^2=x^2+y^2$ is the cylindrical coordinate radius of the potential in the disc plane, and $z$ is the vertical distance over the disk component. The parameters $G$, $M_d$, $A$ and $B$ denote the gravitational constant, the disc mass and the shape of the disc, respectively. Taking $A=0$ the potential becomes the Plummer potential.
The bar is modelled by an ellipsoid with density function 
\begin{equation}\label{eqn:Ferrers}
 \rho = 
 \left\lbrace
 \begin{array}{ll}
  \rho_0(1-m^2)^{n_h}, & m\leq 1 \\
  0,   & m > 1 \\
 \end{array}
 \right.
\end{equation}
where $m^2=x^2/a^2 + y^2/b^2 + z^2/c^2$, \emph{a} (semi-major axis), \emph{b} (intermediate axis) and \emph{c}  (semi-minor axis) determine the shape of the bar, $n_h$ is the homogeneity degree of
the mass distribution ($n_h=2$ in our work) and $\rho_0$ is the density at the origin ($\rho_0=\frac{105}{32\pi}\frac{GM_b}{abc}$ if $n_h=2$, where $M_b$ is the bar mass).

The unit of length considered is the kpc, the time unit is u$_t = 2 \times 10^6$~yr,  $\Omega$ is in~[u$_t$]$^{-1}$, and the mass unit is u$_\text{\tiny{M}} = 2 \times 10^{11} M_\odot$, where $M_\odot$ denotes the mass of the Sun. $G$ stands for the gravitational constant. 

In our model, we select a disc radius of $A=3$~kpc, height $B=1$~kpc and mass giving a value of $GM_d=0.52$~kpc$^3$/$u_t^2$. The dimensions of the bar are $a=6$~kpc, $b=1.5$~kpc, $c=0.4$~kpc, and its mass is such that $GM_b=0.4$~kpc$^3$/$u_t^2$. The Plummer bulge with a radius $B = 1$~kpc and $GM_{bl}$ is set to have around $15\%$ of the mass of the bar $GM_b$, in order that $G(M_d+M_b+M_{bl}) = 1$. The bar pattern speed is fixed as $\Omega=0.0633$~[u$_t$]$^{-1}$ ($\sim 30.97$~km/s/kpc).

In the rotating reference frame aligned with the main axis of the bar, the equations of motion given in Eq.\eqref{eqn:motion} are written as the following dynamical system:
\begin{equation}
\left\lbrace
\begin{array}{l}
 \ddot{x} = 2\,\Omega\, \dot{y} + \Omega^2\, x  - \phi_{x} \\
 \ddot{y} = -2\,\Omega\, \dot{x} + \Omega^2\, y - \phi_{y} \\
 \ddot{z} = - \phi_{z}\,.\\
\end{array}
\right.
\label{eqn:systmodel}
\end{equation}

The Jacobi first integral of Eq.\eqref{eqn:motion} (which can be regarded as the energy in the rotating frame) is given by
 \begin{equation}\label{eqn:PreCJAC}
   C_J(x,y,z,\dot{x},\dot{y},\dot{z}) = -\,(\dot{x}^2+\dot{y}^2+\dot{z}^2)+\Omega^2\,(x^2+y^2)-2\,\phi,
\end{equation}
and the effective potential
is defined by $\phi_{_{\hbox{\scriptsize eff}}} = \phi - \frac{1}{2}\,\Omega^2\, (x^2 + y^2)$.

\begin{figure*}
  \centering
    \includegraphics[width=0.245\textwidth]{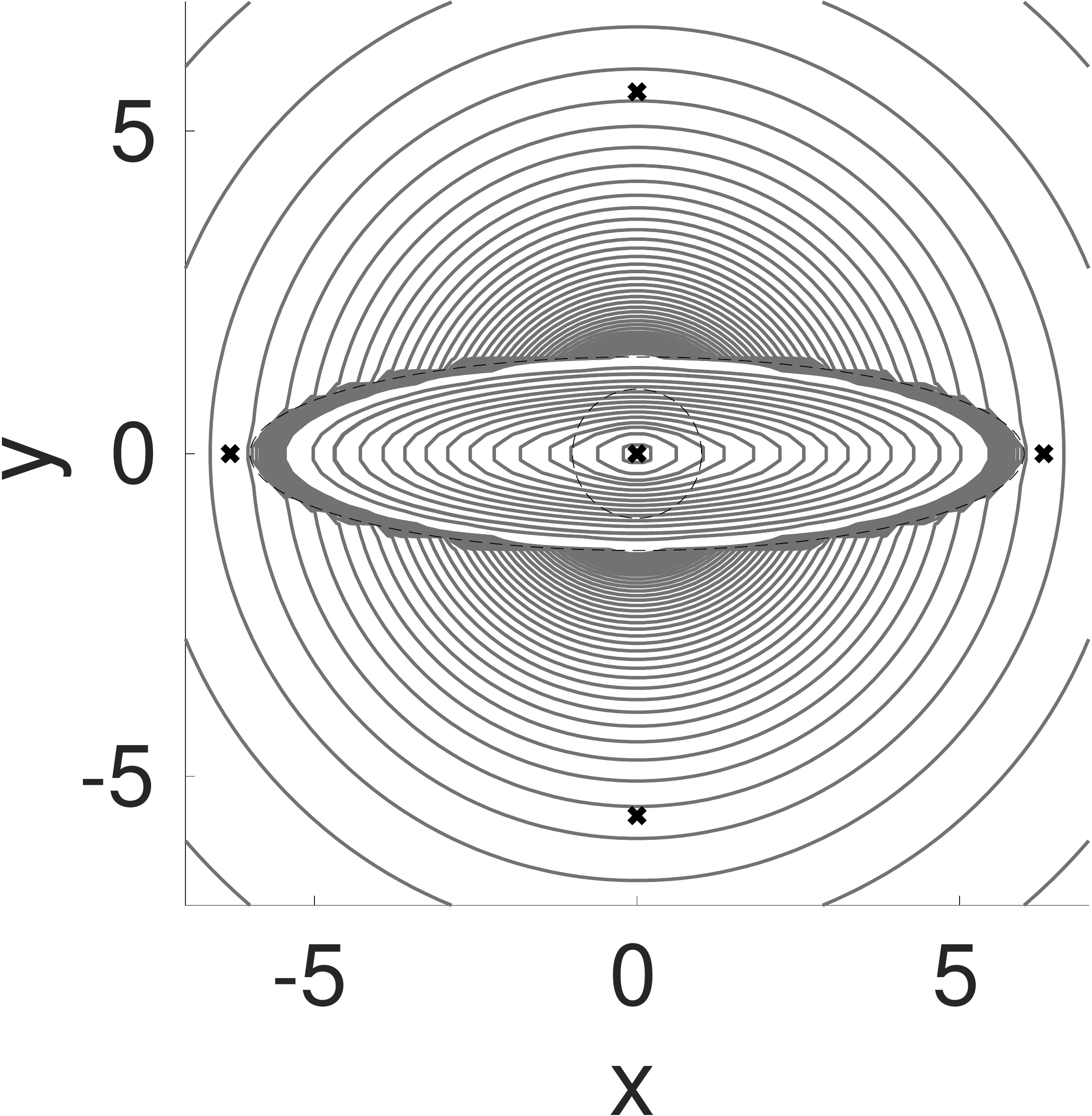}
    \includegraphics[width=0.245\textwidth]{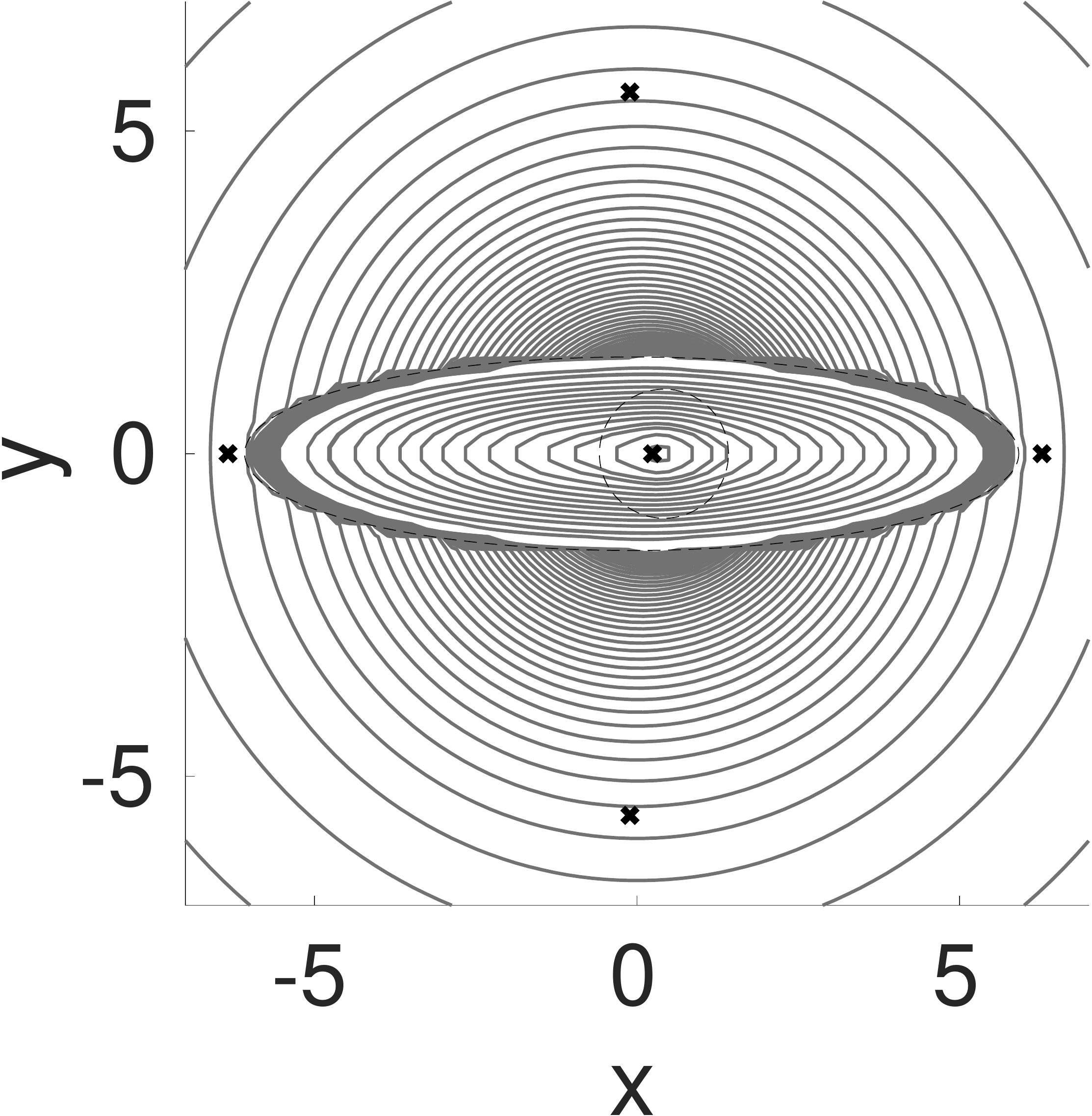}
    \includegraphics[width=0.245\textwidth]{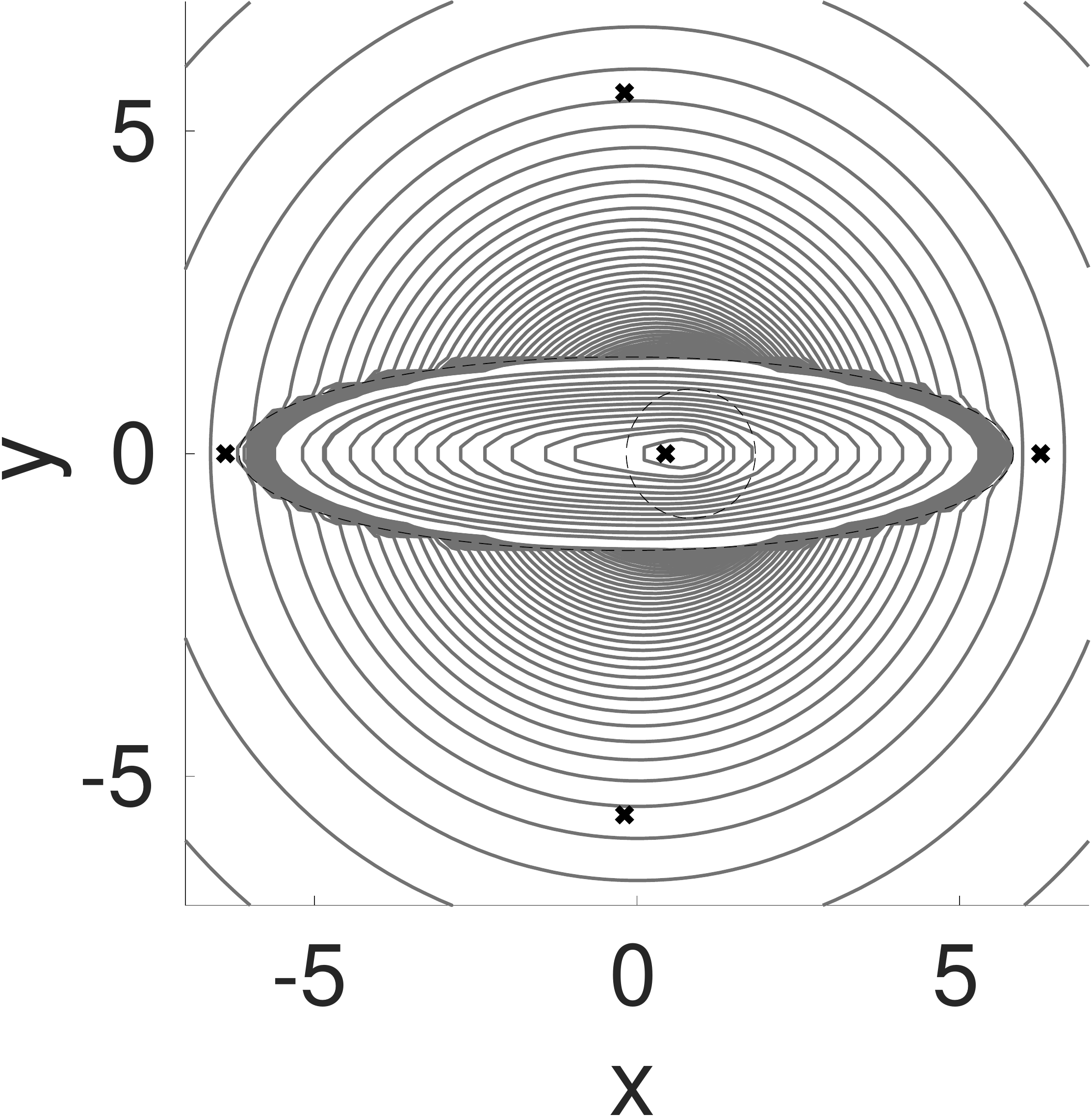}
    \includegraphics[width=0.245\textwidth]{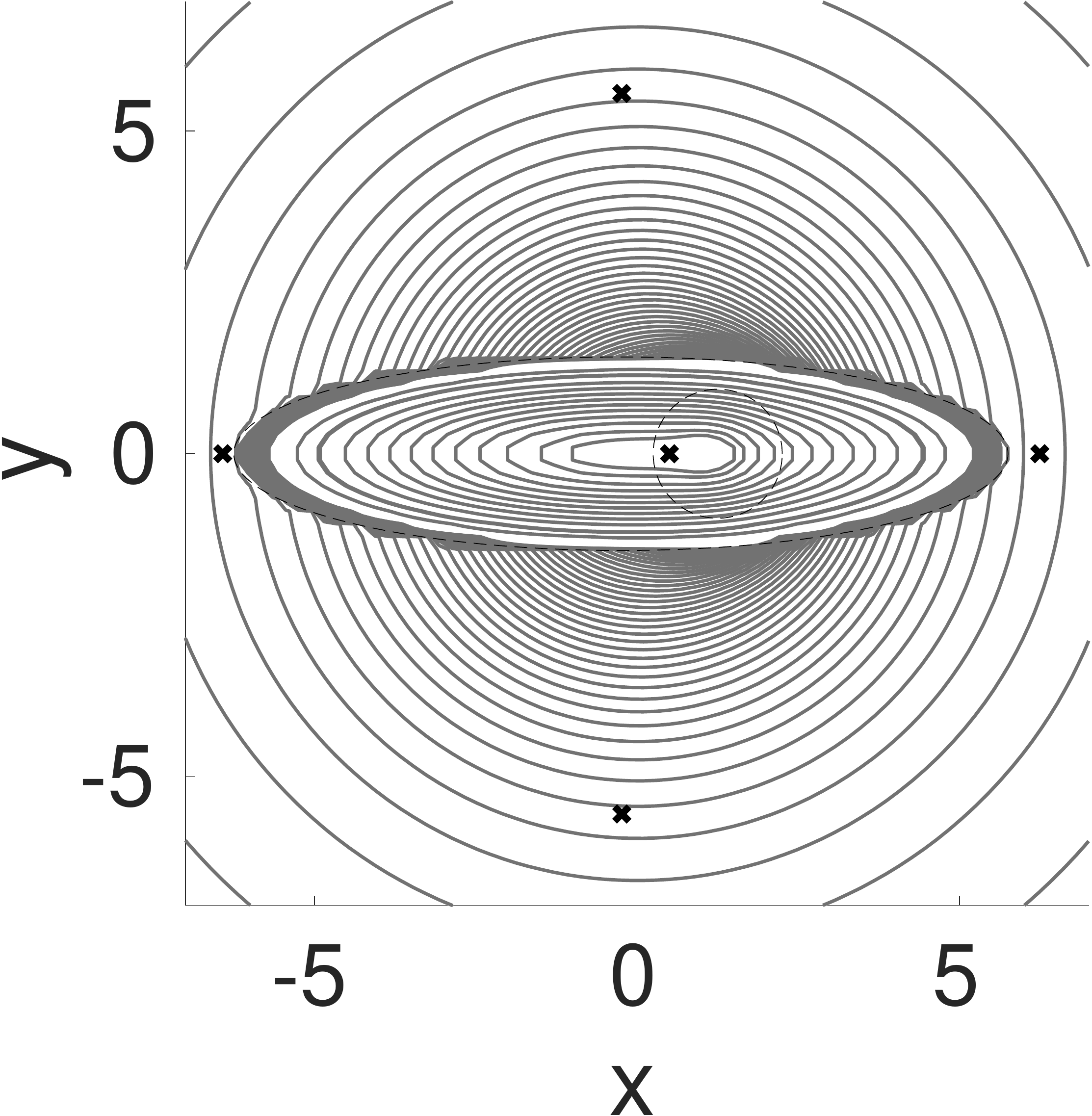}
  \caption{Isodensity curves of the potential $\phi = \phi_d+\phi_b+\phi_{bl}$. Equilibrium points of the system are marked with a cross. The Ferrers bar and the Plummer bulge
are outlined by dotted black curves. From left to right: Bulge centered at $(0,0,0)$, $(0.5,0,0)$, $(1,0,0)$ and at $(1.5,0,0)$.}
  \label{fig:isodensity}
\end{figure*}

The goal of this paper is to analyze which is the effect of an asymmetric  distribution of mass in the central parts of the galaxy on the external spiral structure. To introduce this asymmetry, we displace the bulge potential along the $x$-axis (main axis of the bar) towards the equilibrium point placed at the right end of the bar. The bulge center is then located at $(x_d,0,0)$ using several values for the displacement $x_d$ (in kpc), namely: $(0,0,0)$, $(0.5,0,0)$, $(1,0,0)$ and $(1.5,0,0)$. We move the center of our potential model to the resulting center of mass of the system. The plot of the equal density contours of the resulting models are shown in Fig.~\ref{fig:isodensity}. Note that the displacement of the bulge along the main axis of the bar creates an asymmetry in the central part of the model and around the libration points $L_1$ and $L_2$. 

\begin{figure}
  \centering
    \includegraphics[width=0.4\textwidth]{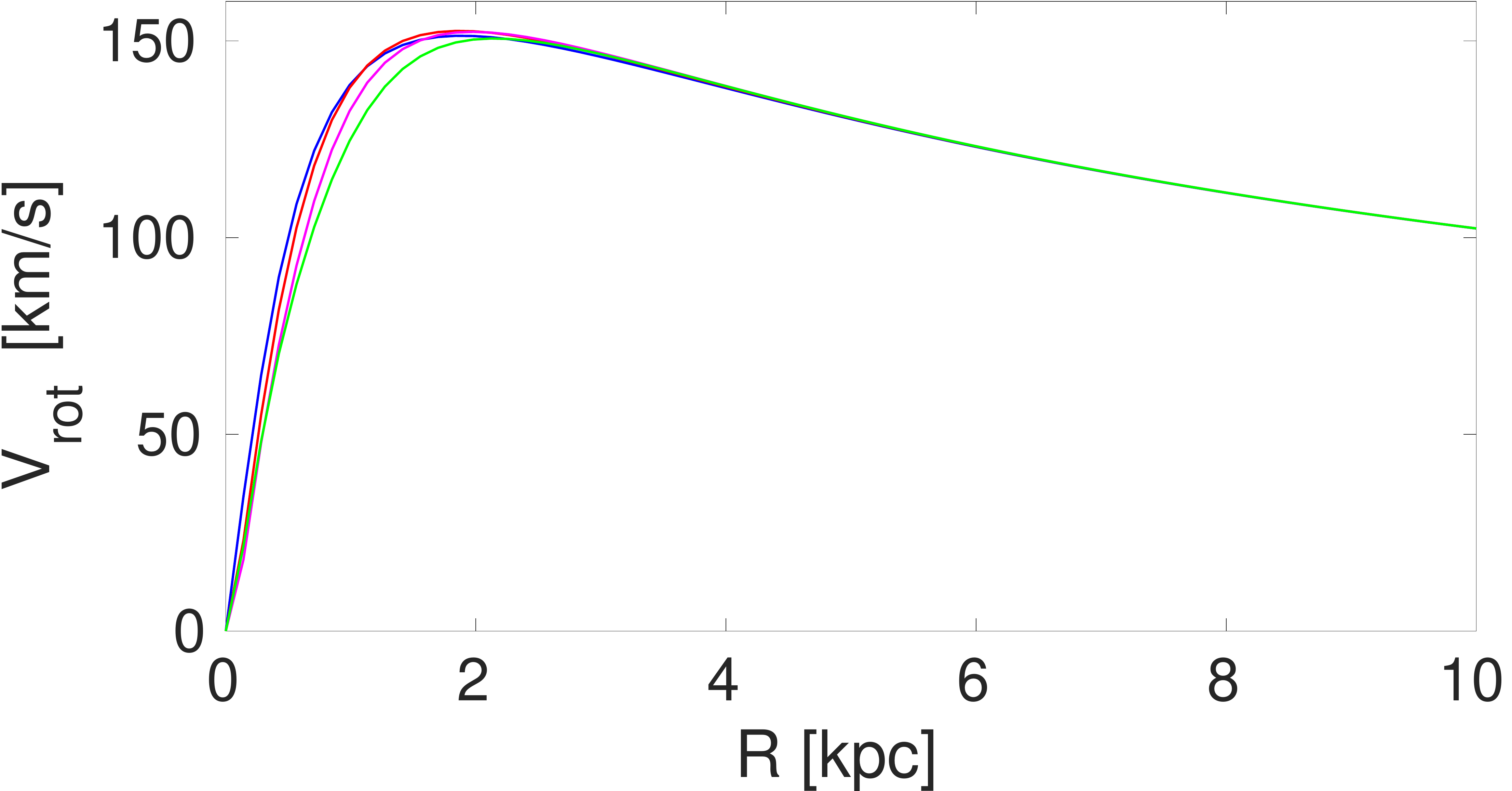}
  \caption{Rotation curve of the potential $\phi = \phi_d+\phi_b+\phi_{bl}$ for the bulge centered at $(0,0,0)$ (in blue), $(0.5,0,0)$ (in red), $(1,0,0)$ (in magenta) and $(1.5,0,0)$ (in green).}
  \label{fig:rotcurve}
\end{figure}

The rotation curves of the model, defined as $V_{\text{rot}}^2 = r\dfrac{d\phi}{dr}$,
where the potential is $\phi = \phi_d+\phi_b+\phi_{bl}$ with the selected values of the parameters and for the above explained positions of the  bulge displacement, are shown in Fig.~\ref{fig:rotcurve}.
The resulting rotation curve is reasonably flat in the outer parts, and displays only minor differences in the position of the maximum.


\section{Dynamics of the models}
\label{sec:dynamics}

The solutions of $\nabla\phi_{_{\hbox{\scriptsize eff}}} = 0$ in rotating coordinates give five Lagrangian equilibrium points of the model ($L_i$, $i=1,\ldots,5$). Points $L_1$ and $L_2$ are linearly unstable points and lie on the $x$-axis at the ends of the bar. Point $L_3$ is linearly stable and it is placed on the origin of coordinates in the case of a symmetric model. Points $L_4$ and $L_5$ are also linearly stable and located out of the $x$-axis. A detailed explanation of the dynamics around these points can be found in \citet{Athan1983, Romero1, Warps}.

\begin{figure*}
  \centering
    \includegraphics[width=0.245\textwidth]{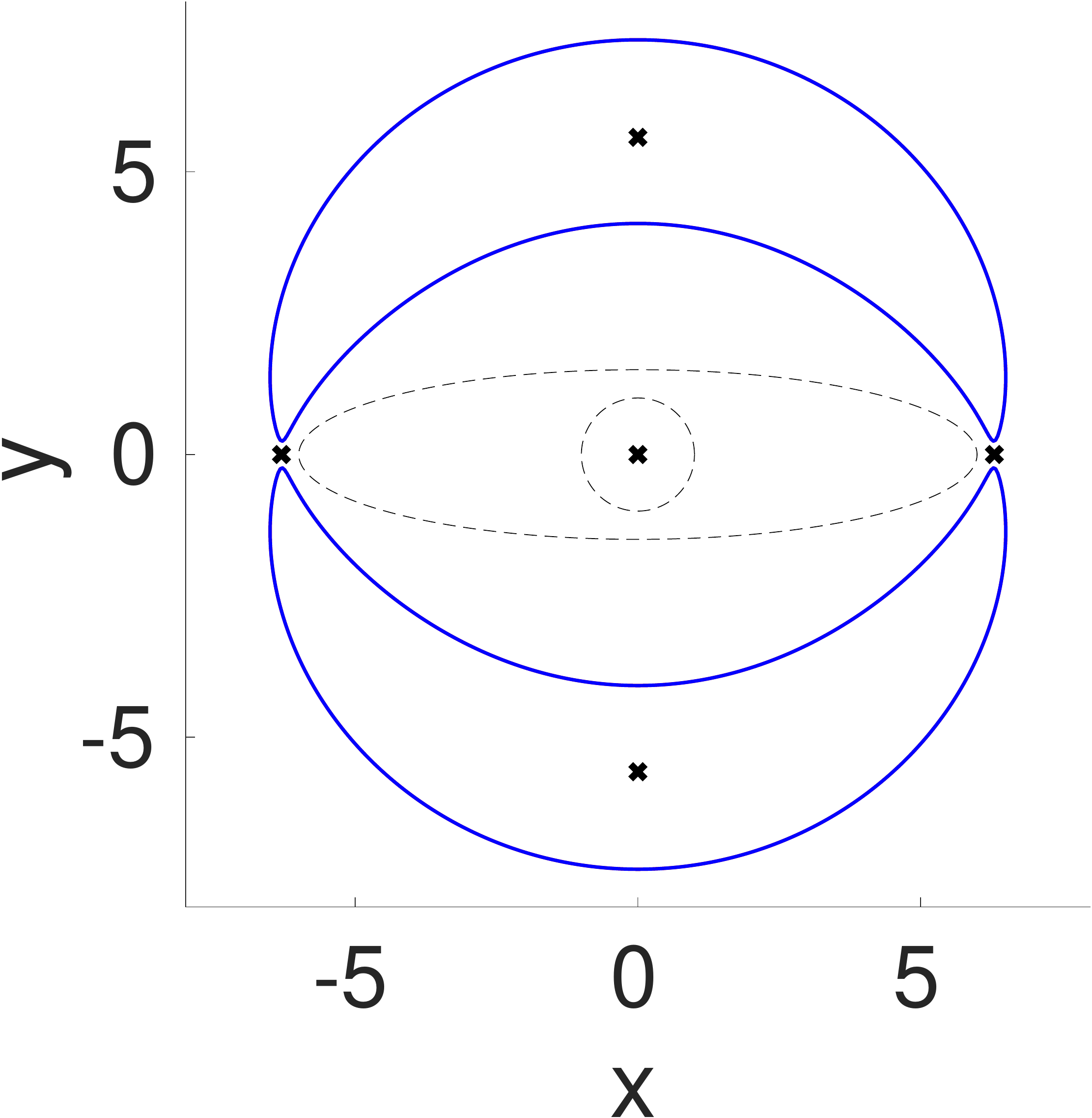}
    \includegraphics[width=0.245\textwidth]{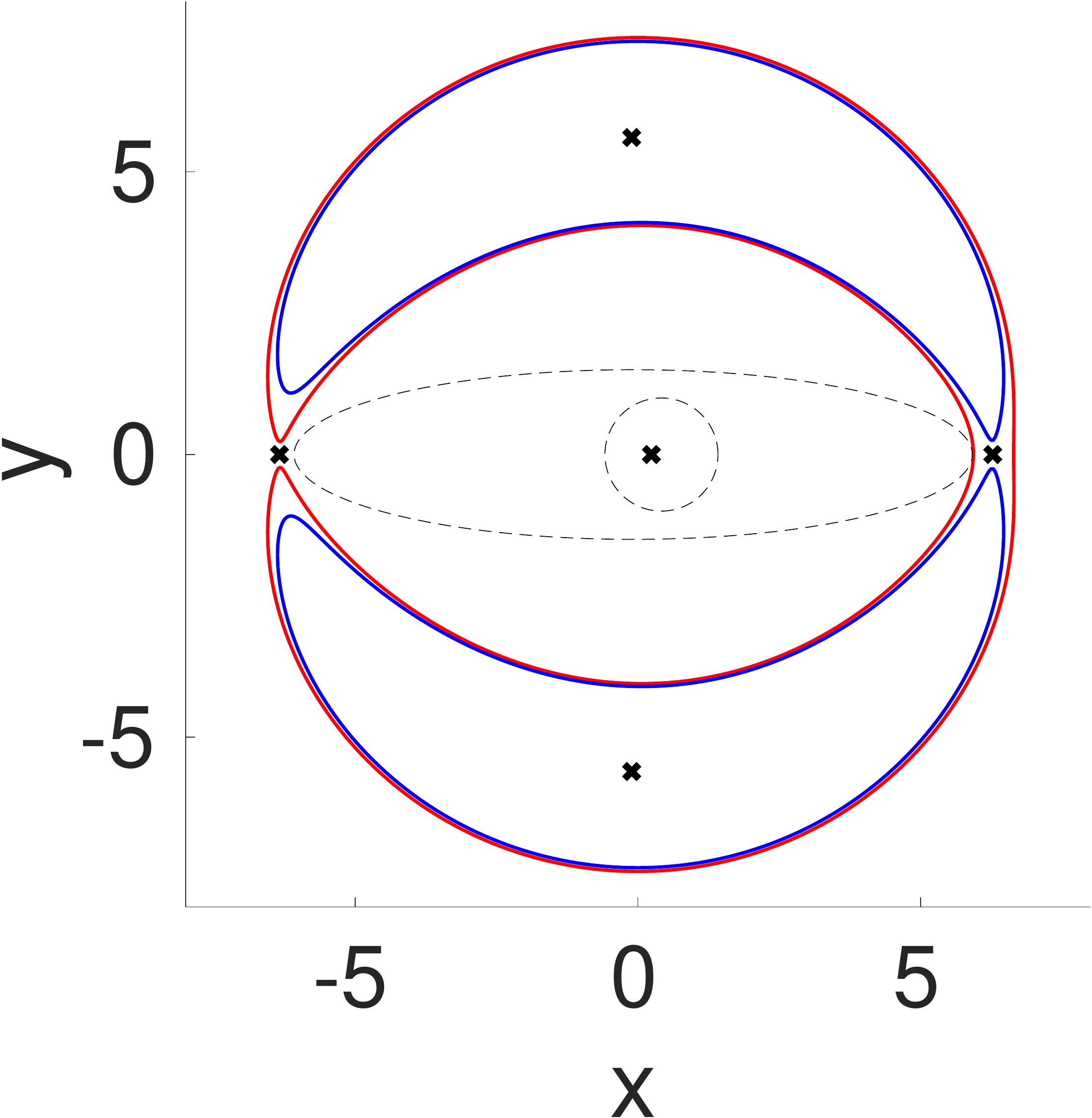}
    \includegraphics[width=0.245\textwidth]{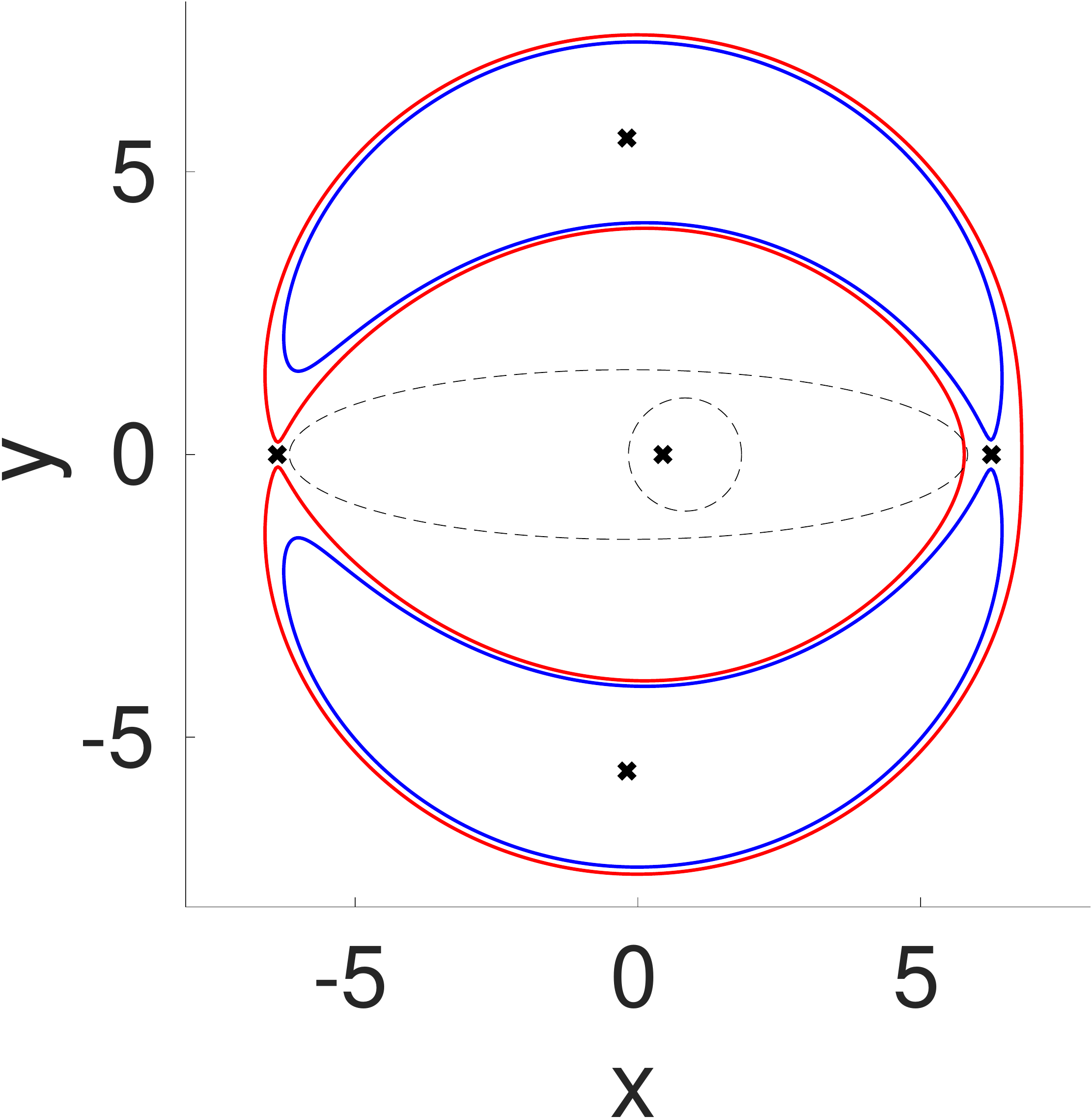}
    \includegraphics[width=0.245\textwidth]{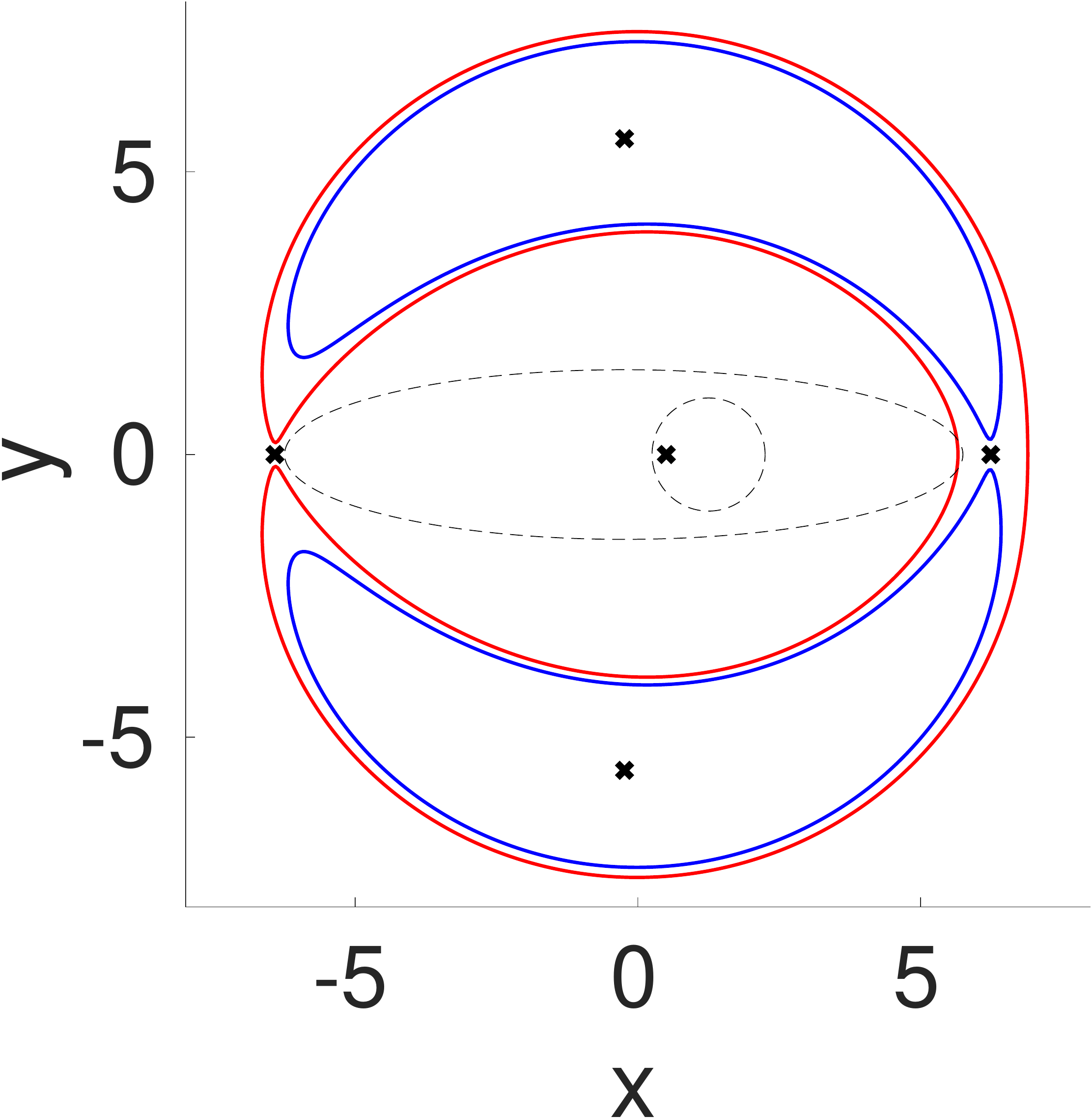}
  \caption{Zero velocity curves for a Jacobi constant slightly above to that of $L_1$ (blue) and for one slightly above to that of $L_2$ (red). Equilibrium points of the system are marked with a cross. The Ferrers bar and the Plummer bulge are outlined by dotted black curves. From left to right: Bulge centered at $(0,0,0)$, $(0.5,0,0)$, $(1,0,0)$ and at $(1.5,0,0)$.}
  \label{fig:cvz}
\end{figure*}

The regions where $\phi_{_{\hbox{\scriptsize eff}}}>C_J$ are forbidden regions for a star of energy $C_J$. In the plane, these regions are delimited by the zero velocity curves, which are defined by the level surfaces $\phi_{_{\hbox{\scriptsize eff}}}=C_J$ intersected with $z=0$. In Fig.~\ref{fig:cvz} we show the zero velocity curves corresponding to an energy slightly above of that of the equilibrium point, $C_{J,{L_i}}+\delta$, for the models with off-centered bulges in Fig.~\ref{fig:rotcurve}, i.e., setting the values $x_d = 0,\,0.5,\,1,\,1.5$. These zero velocity curves limit the inner and outer regions in the galaxy. In the symmetric case ($x_d = 0$), the energy of both equilibrium points $L_1$ and $L_2$ is the same, $C_{J,{L_1}} = C_{J,{L_2}}$. For the asymmetric cases, as $x_d$ grows $C_{J,{L_2}}$ becomes smaller than $C_{J,{L_1}}$, which makes the zero velocity curves related to $L_1$ more open at the opposite point. 

\begin{figure}
  \centering
    \includegraphics[width=0.237\textwidth]{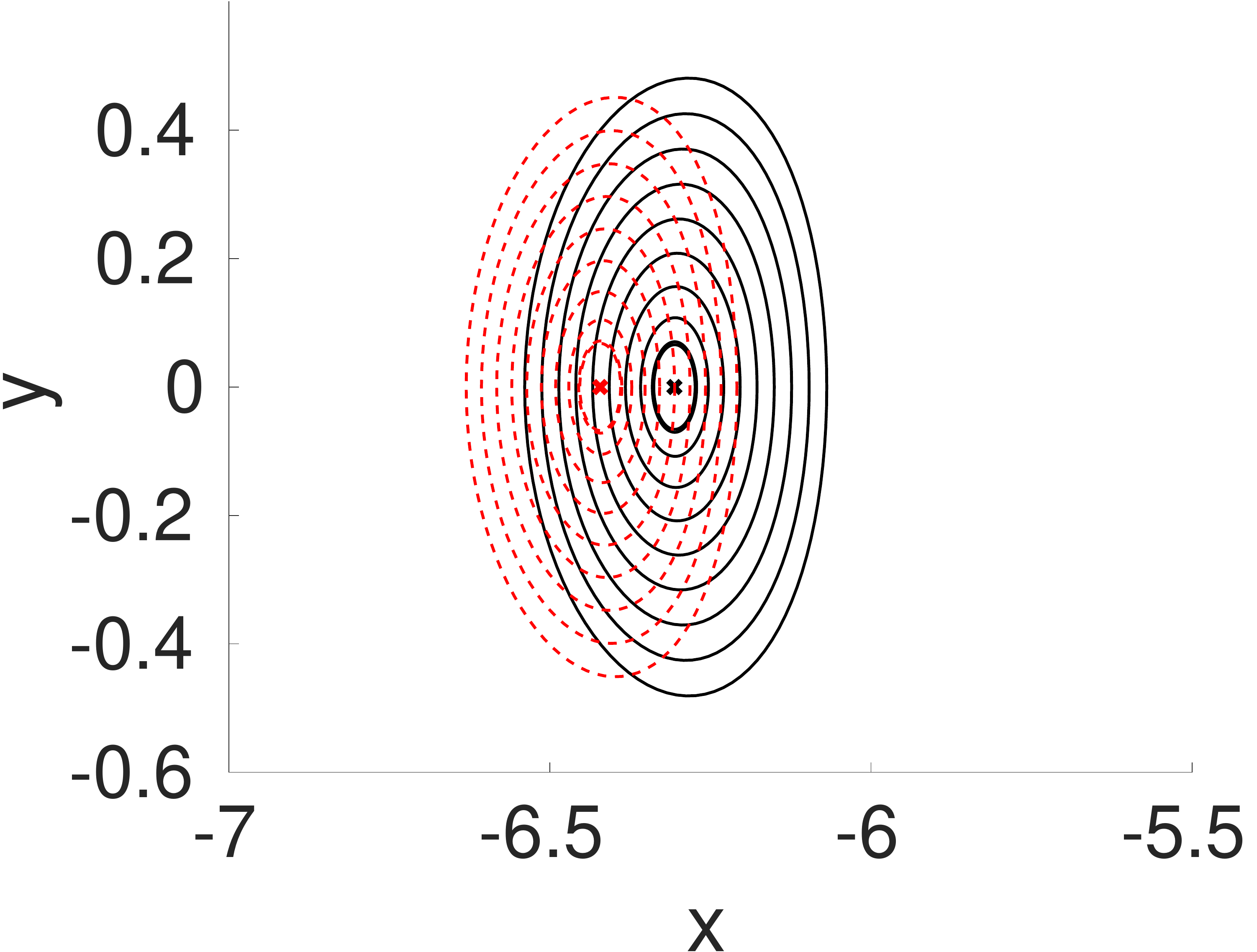}
    \includegraphics[width=0.237\textwidth]{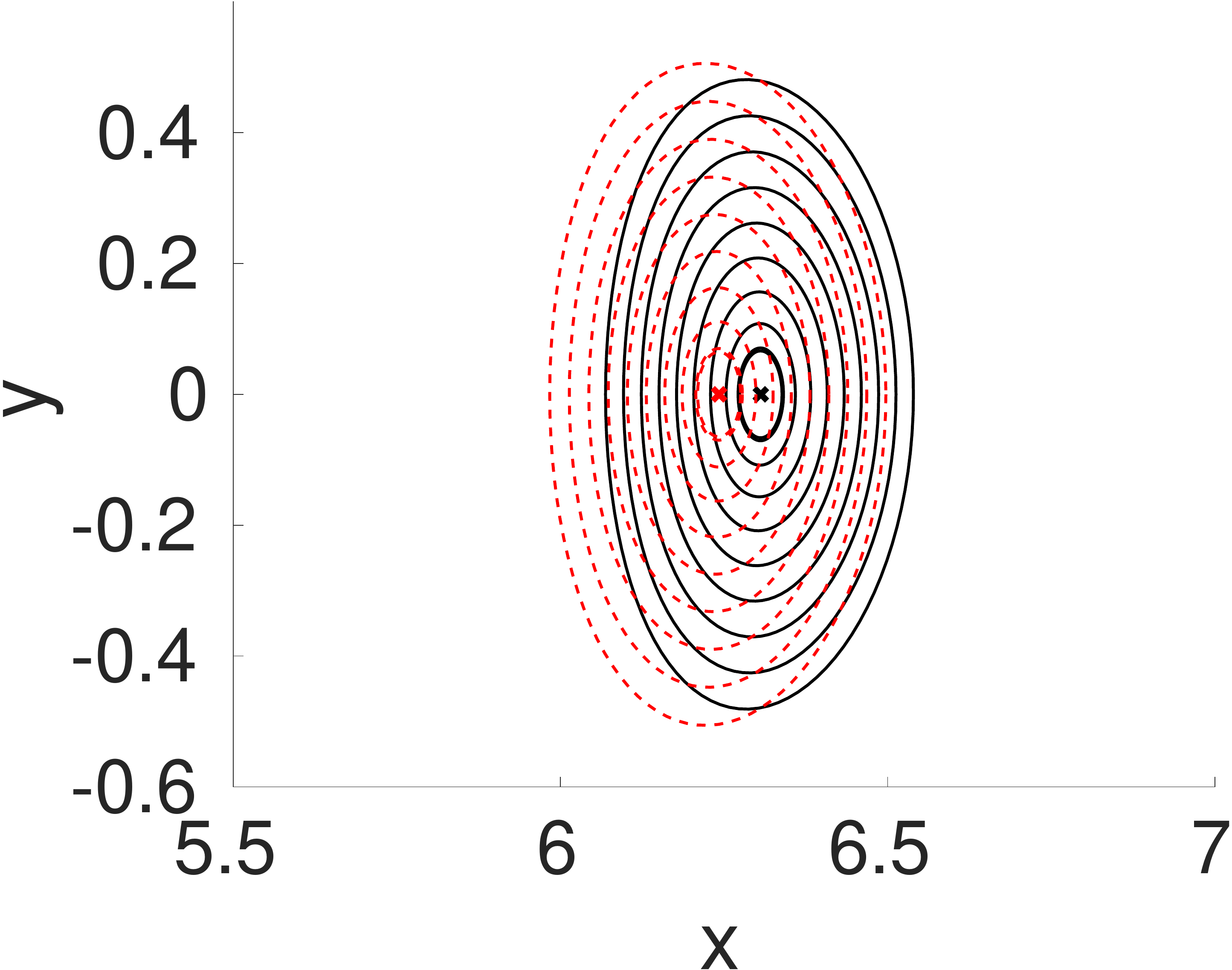}
  \caption{Lyapunov family of periodic orbits around $L_i$, $i=1,\,2$, for a range of values of the Jacobi constant in $(C_{J,{L_i}},\, C_{J,{L_i}} + 10^{-4})$. Solid black line: $(x,\,y)$ projection of the family around $L_2$ (left) and $L_1$ (right) for the symmetric model with bulge centered at $(0,0,0)$, $C_{J,{L_i}} = -0.2338$, $i=1,\,2$. Dotted red line: $(x,\,y)$ projection of the family around $L_2$ (left) and $L_1$ (right) for the asymmetric model with bulge centered at $(1.5,0,0)$, $C_{J,{L_2}} = -0.2358$, $C_{J,{L_1}} = -0.2331$.}
  \label{fig:Lyap}
\end{figure}

Particular attention is given in this work to the unstable points $L_1$ and $L_2$. They are surrounded by planar and vertical families of Lyapunov periodic orbits, which are unstable in the neighborhood of the equilibrium point. The relevant family for the transport of matter between the inner and outer regions of the galaxy is the planar family \citep{Rom09}. Fig.~\ref{fig:Lyap} shows the $(x,\,y)$ projection of the planar family for the models with $x_d=0$ (solid black line) and $x_d=1.5$ (dotted red line). The family around $L_2$ is displayed at the left panel and that around $L_1$ at the right one. In the symmetric case ($x_d=0$) both families coincide, in the asymmetric one ($x_d=1.5$) the family around $L_2$ is smaller than the one around $L_1$.  

\begin{figure*}
  \centering
    \includegraphics[width=0.245\textwidth]{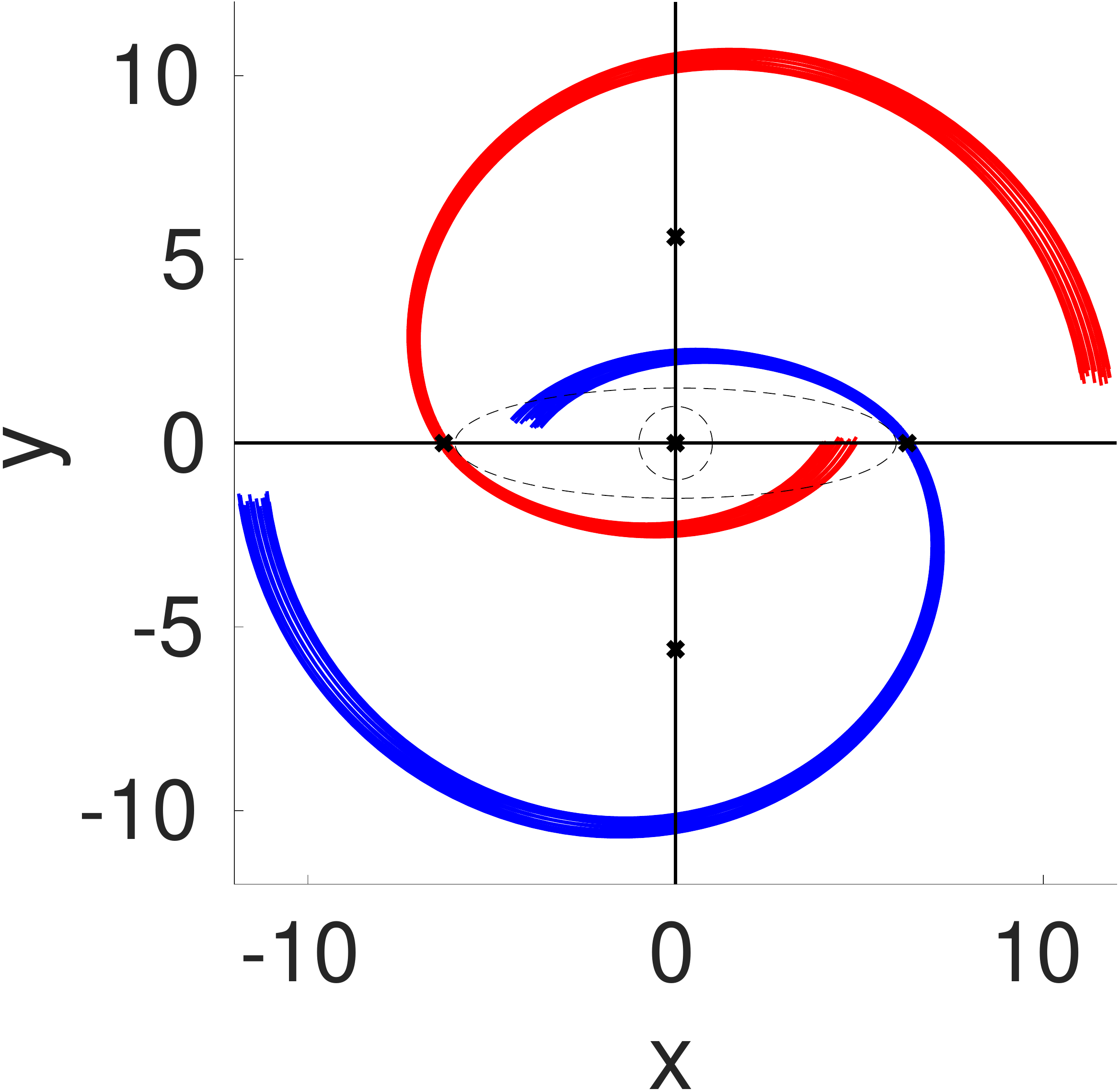}
    \includegraphics[width=0.245\textwidth]{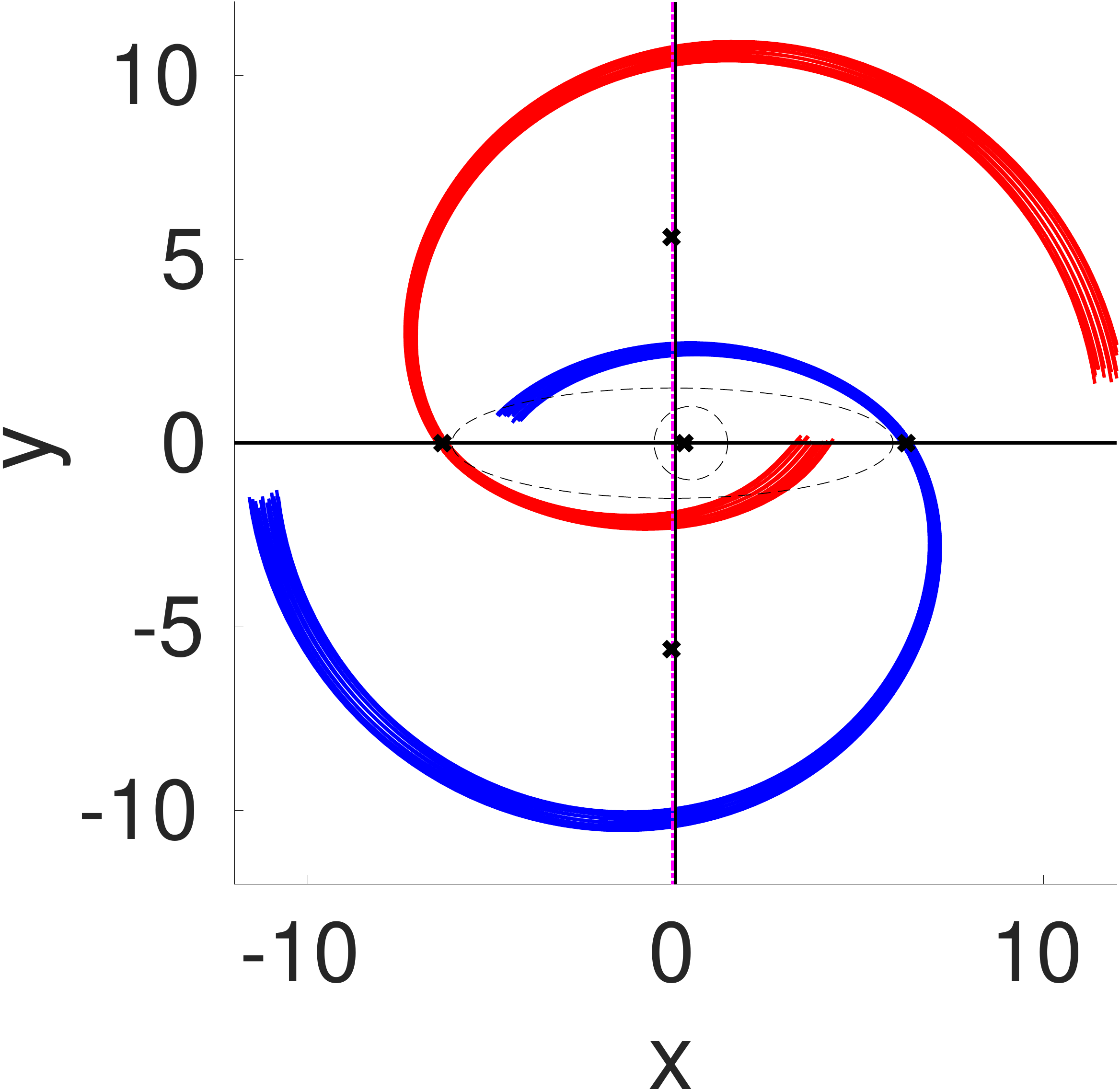}
    \includegraphics[width=0.245\textwidth]{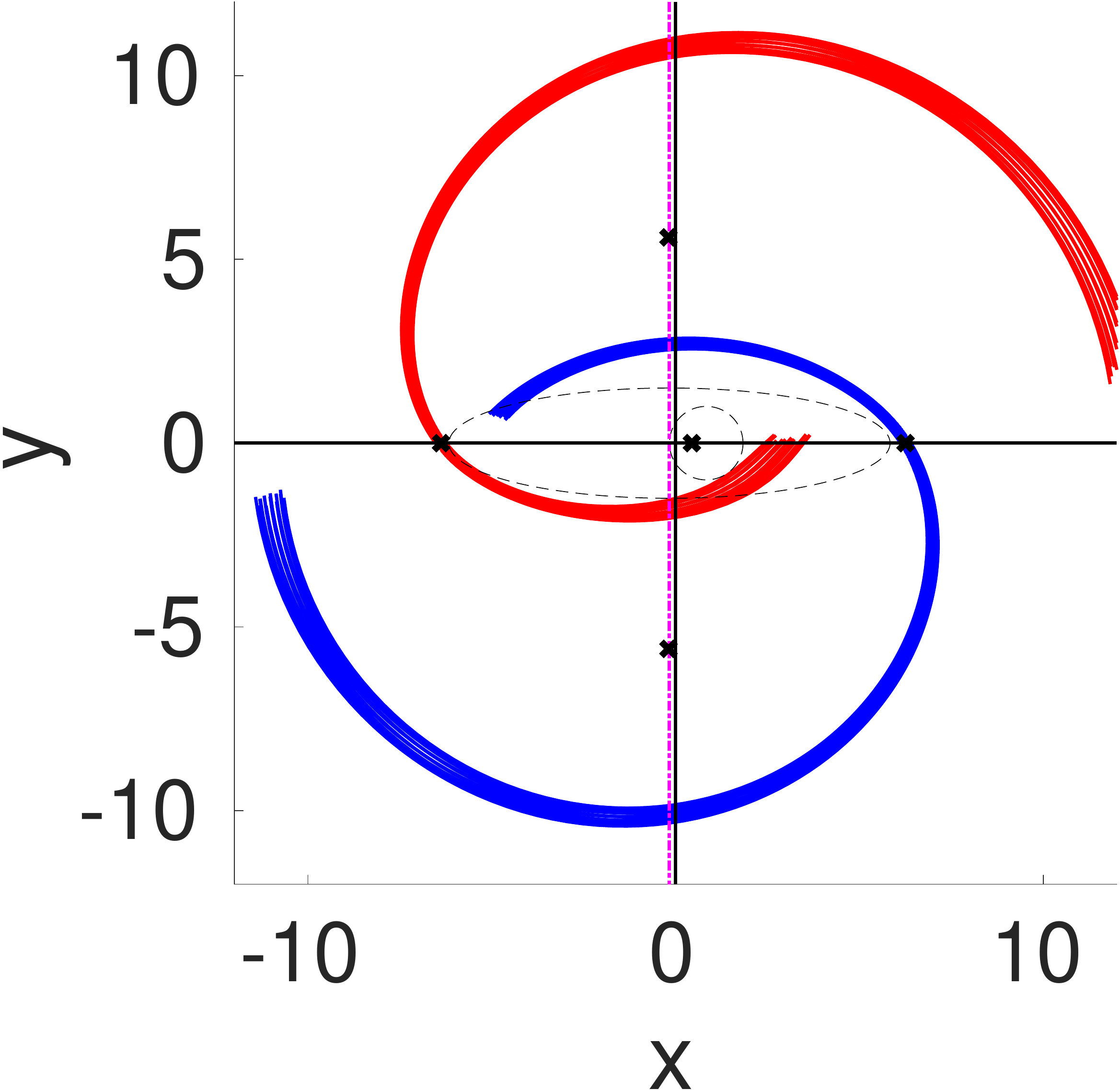}
    \includegraphics[width=0.245\textwidth]{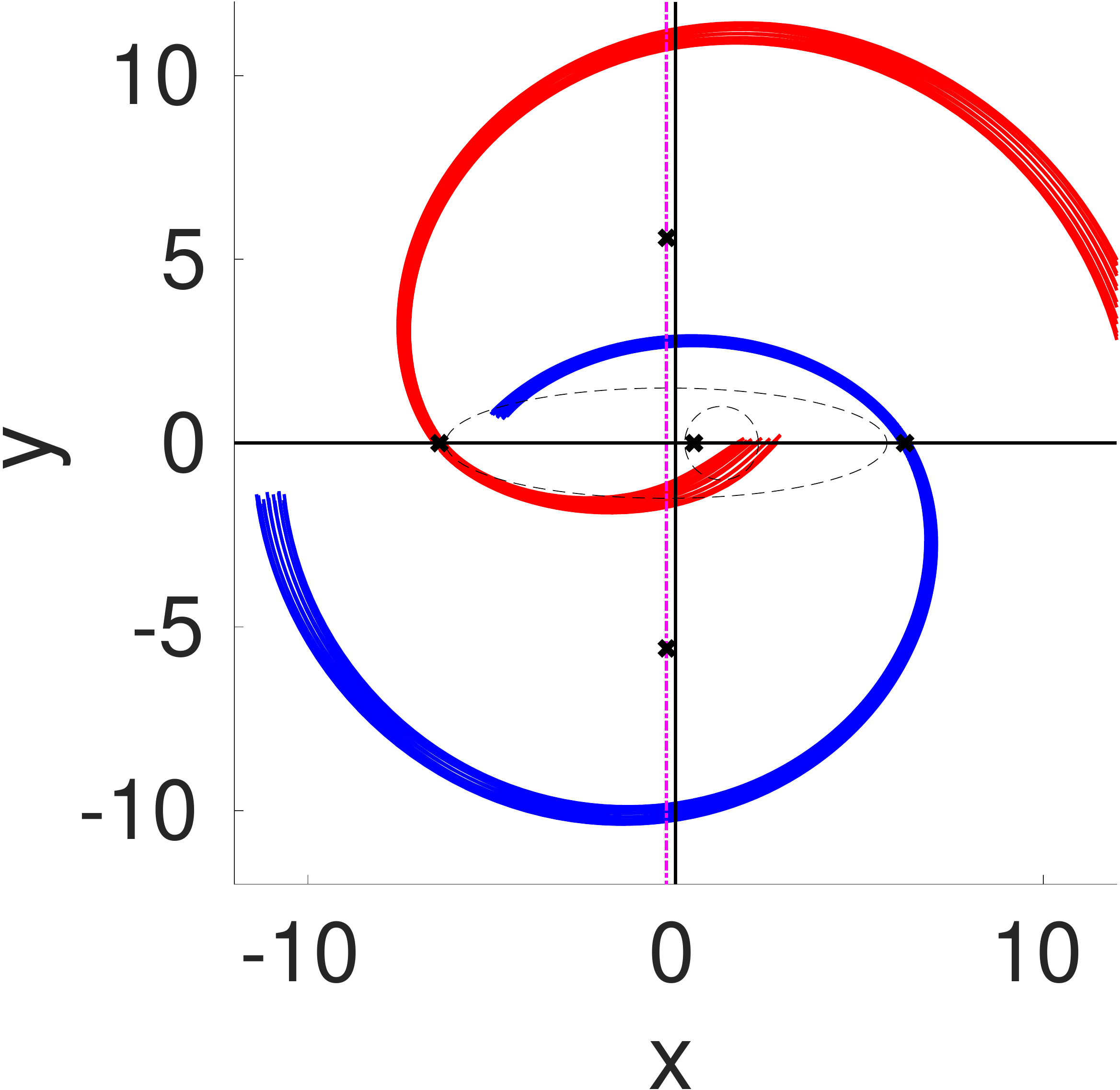}
  \caption{Unstable invariant manifolds associated to the Lyapunov periodic orbits of $L_1$ and $L_2$. The position of the equilibrium points is marked with crosses. The bar and the bulge are outlined by dotted black curves. The reference system is marked with a solid black line and the center of the bar with a dotted magenta line. From left to right: Bulge centered at $(0,0,0)$, $(0.5,0,0)$, $(1,0,0)$ and at $(1.5,0,0)$.}
  \label{fig:manifolds}
\end{figure*}

These critical points are characterized by the superposition of a saddle and two harmonic oscillations in the rotating frame. Consequently, for a given Jacobi constant, stable and unstable invariant manifolds emanate from the periodic Lyapunov orbit around each point. The stable manifold is defined as the set of orbits that asymptotically tend to the periodic orbit forward in time, and the unstable manifold consists of those orbits which depart asymptotically from the periodic orbit. These latter manifold drive the escape orbits that are responsible for the visible trajectories, in the form of arms and rings. Fig.~\ref{fig:manifolds} shows the invariant manifolds associated to $L_1$ and $L_2$ for the set of models where $x_d = 0,\,0.5,\,1,\,1.5$. The effect of an asymmetric mass distribution, modelled by the displacement of the bulge, makes the exterior manifold that emanates from $L_2$ to differ from the invariant manifold associated with $L_1$.   

The transit orbits trapped inside the manifolds are in charge of the transfer of matter, from the inner to the outer regions delimited by the zero velocity curves \citep{GideaMasdem2007, Romero1, LCS2018}. As the dynamics of our system~\eqref{eqn:motion} takes place in a four dimensional phase space when we consider orbits with $z=\dot{z}=0$ (the plane $z=0$ is invariant), the intersection of the trajectories of the inner branch of the stable invariant manifold of a Lyapunov orbit with the hyperplane $S$ defined by the section $x=0$ in phase space gives a closed curve in the $(y,\,\dot{y})$ projection. Given a pair $(y,\dot{y})$ and an energy level, this lets us to define a state on $S$ by selecting $(0,y,0,\dot{x},\dot{y},0)$, where $\dot{x}$ is determined by the fixed energy level of the Lyapunov orbit and the orientation of the crossing, taking into account Eq.~\eqref{eqn:PreCJAC},
\begin{equation}
\dot{x} = \sqrt{-\dot{y}^2+\Omega^2y^2-2\phi(0,y,0)-(C_{J,{L_i}}+\delta)},  
\end{equation}
with $C_{J,{L_i}}+\delta$ the energy of the Lyapunov orbit around the point $L_i$, $i=1,2$, slightly above of that of the equilibrium point. The forward in time integration of initial conditions corresponding to $(y,\dot{y})$ points inside the closed curve establishes the trajectories of the particles confined inside the invariant manifold that transit from the inner to the outer region. 

\begin{figure*}
  \centering
    \includegraphics[width=0.245\textwidth]{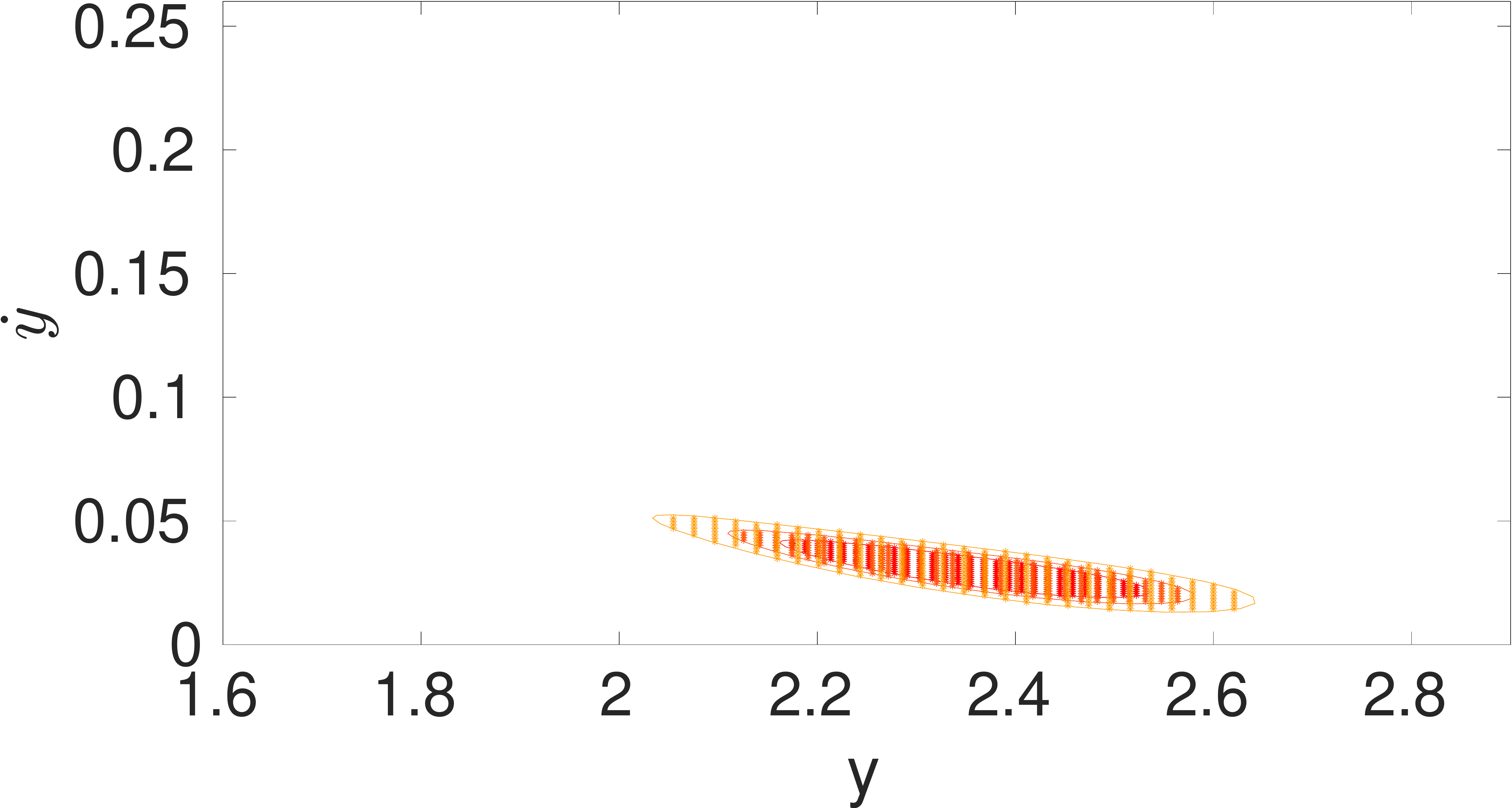}
    \includegraphics[width=0.245\textwidth]{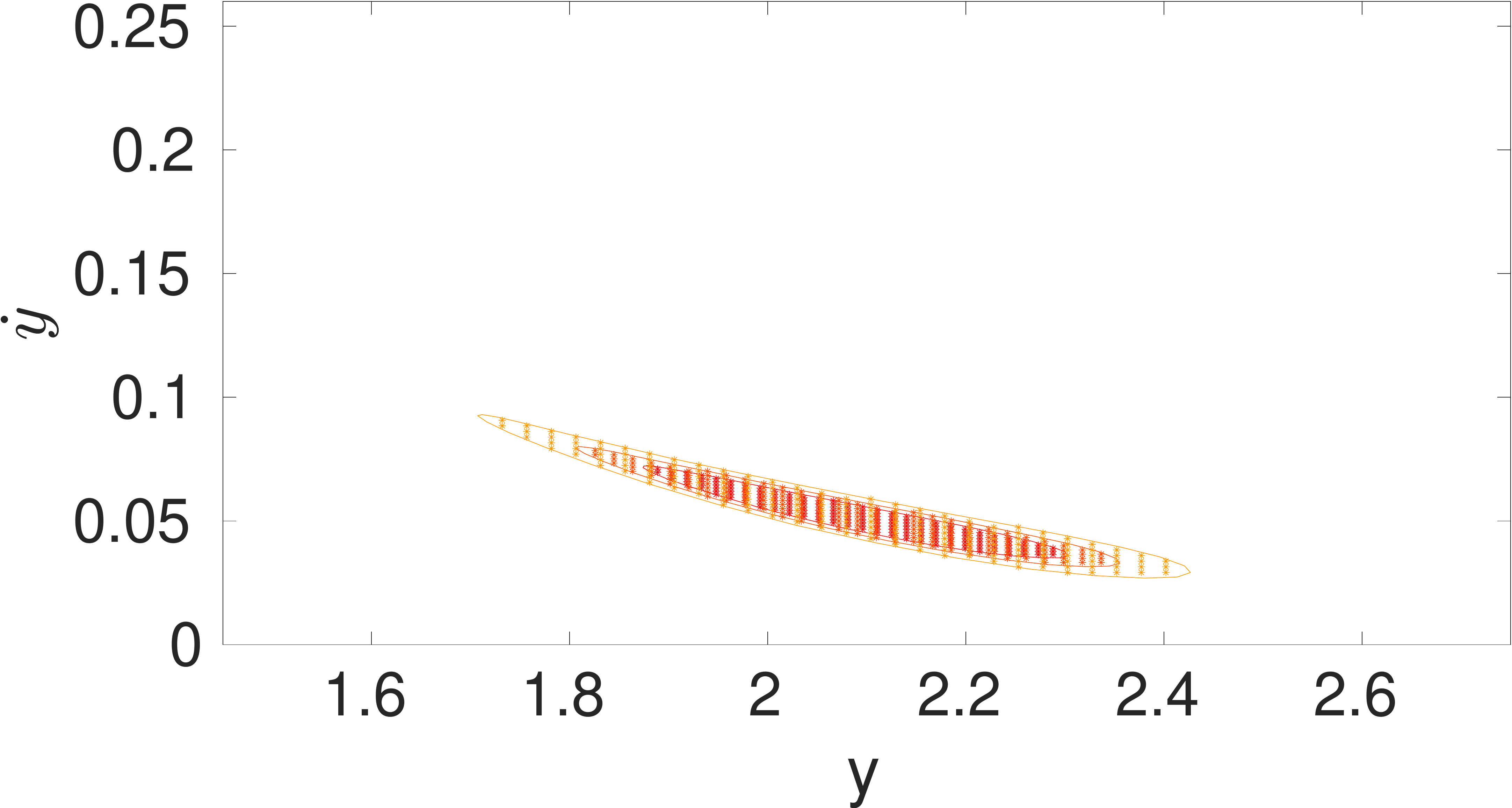}
    \includegraphics[width=0.245\textwidth]{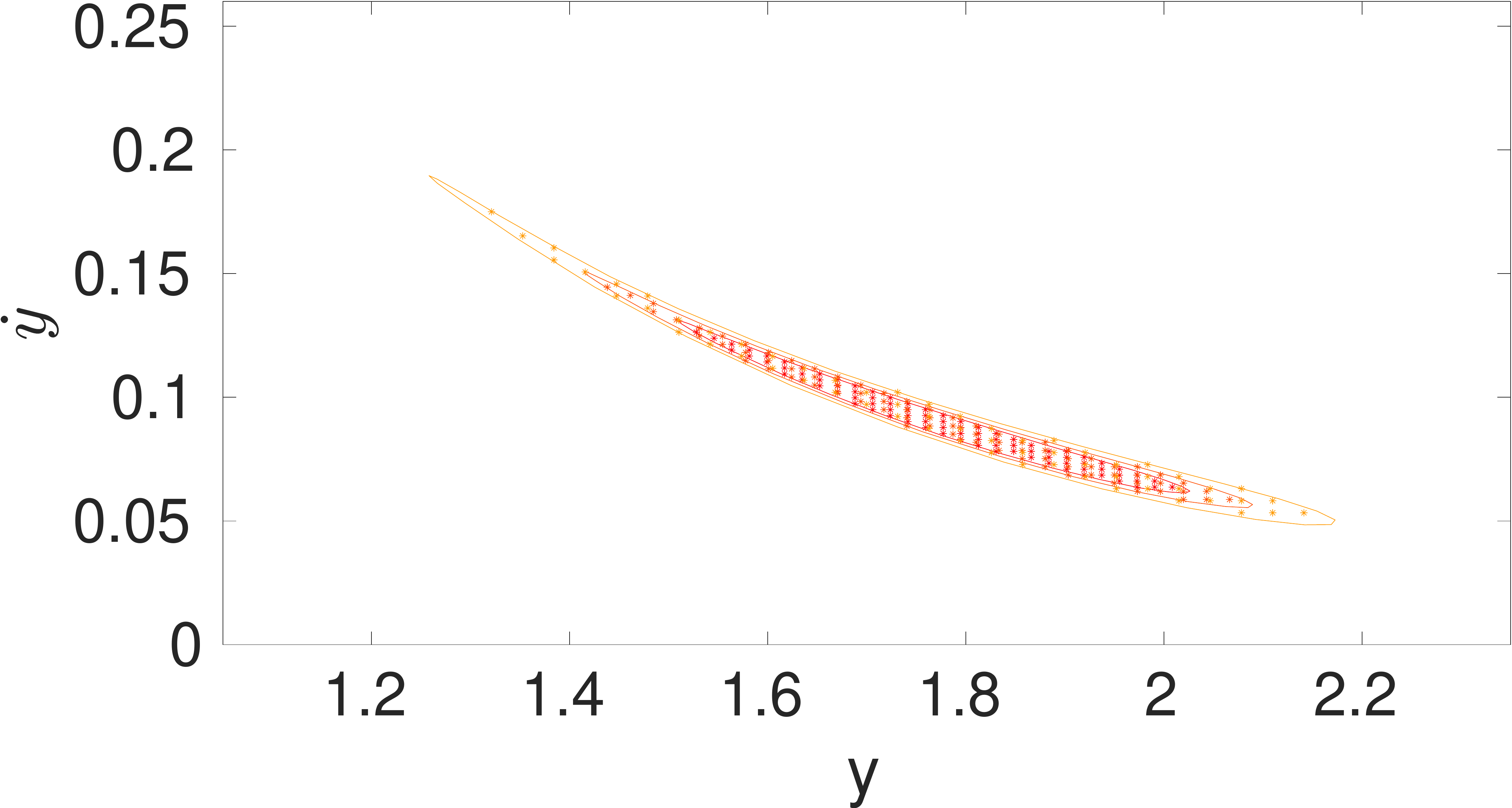}
    \includegraphics[width=0.245\textwidth]{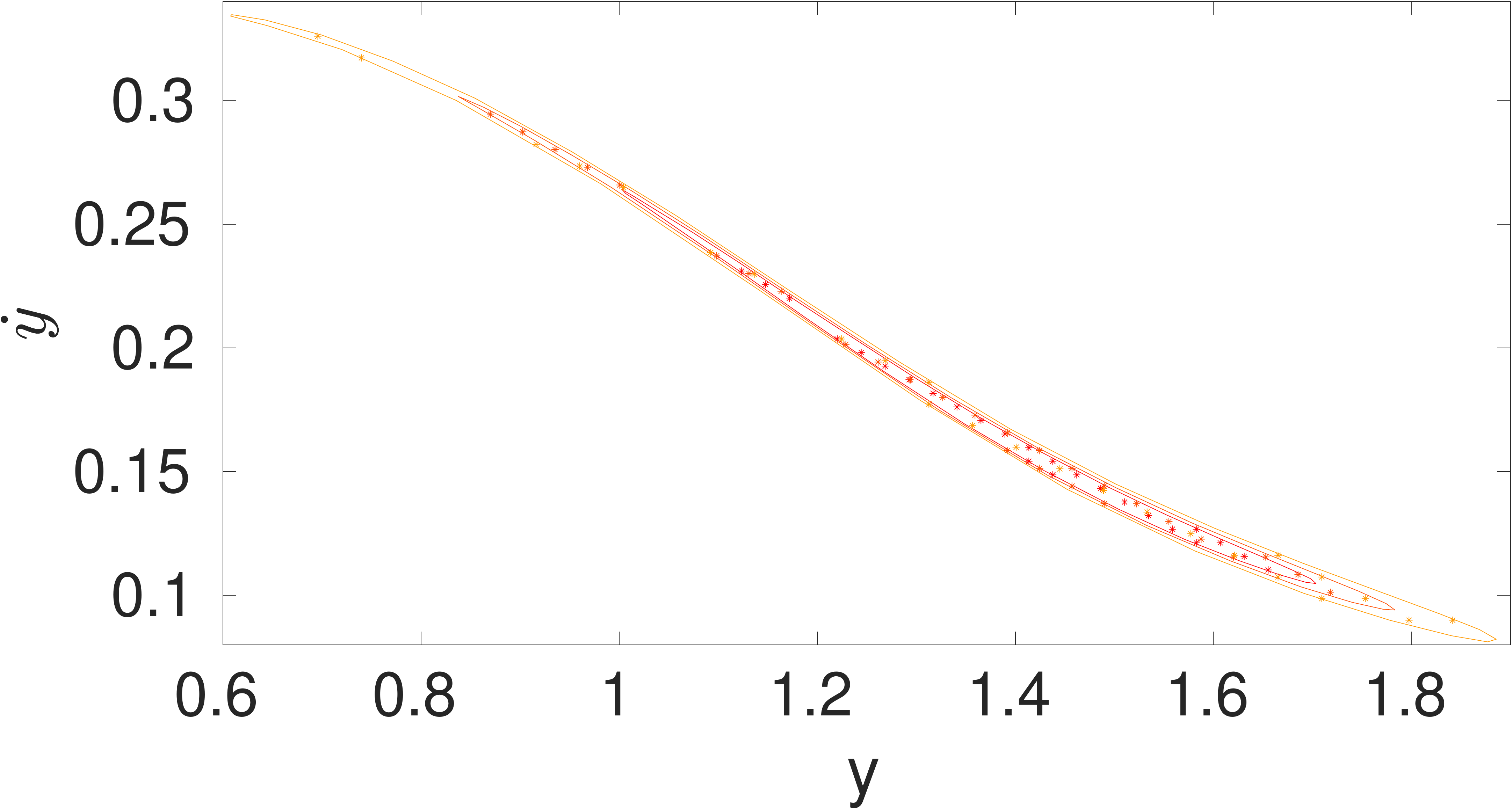}
    
    \includegraphics[width=0.245\textwidth]{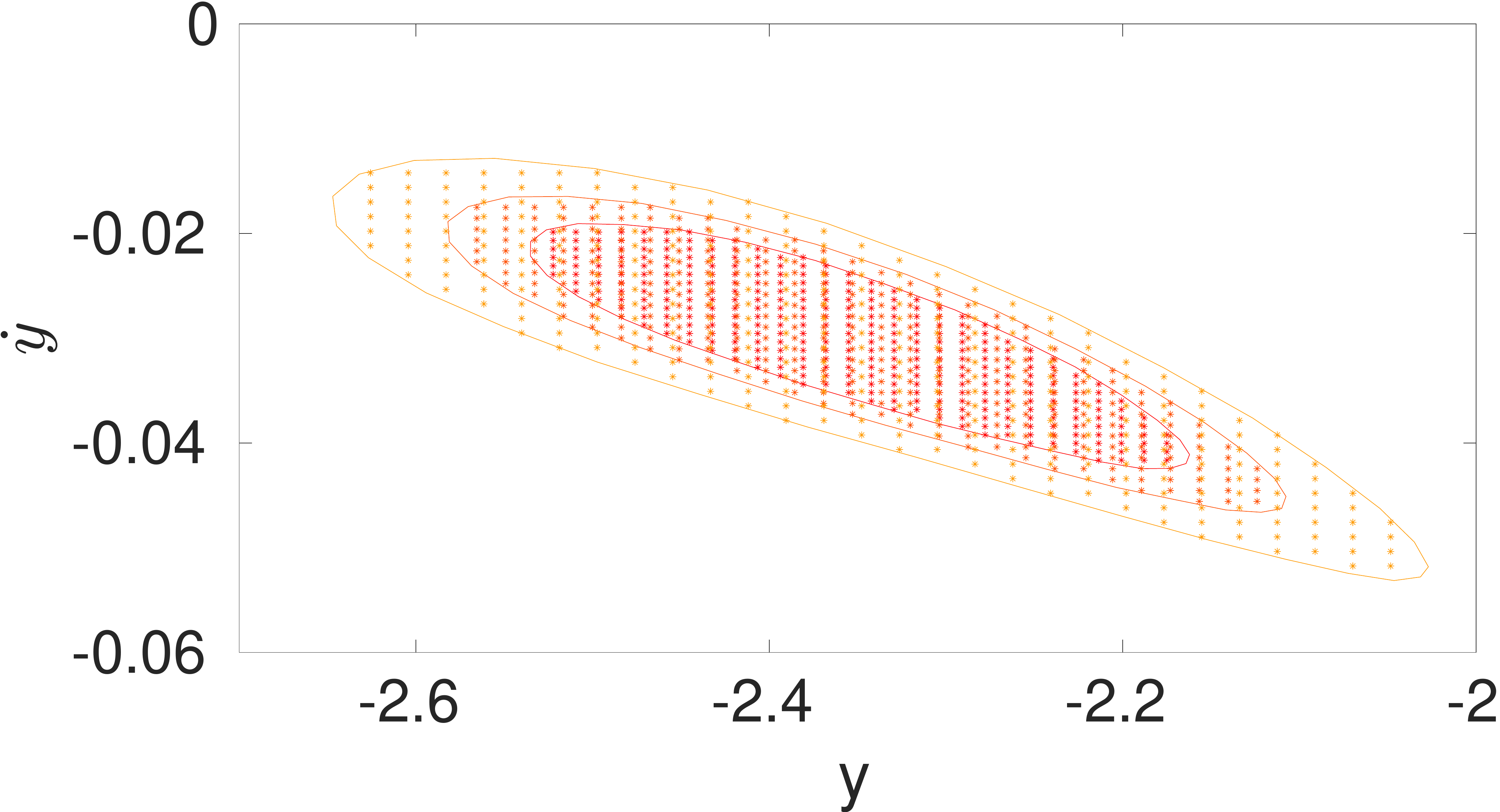}  \includegraphics[width=0.245\textwidth]{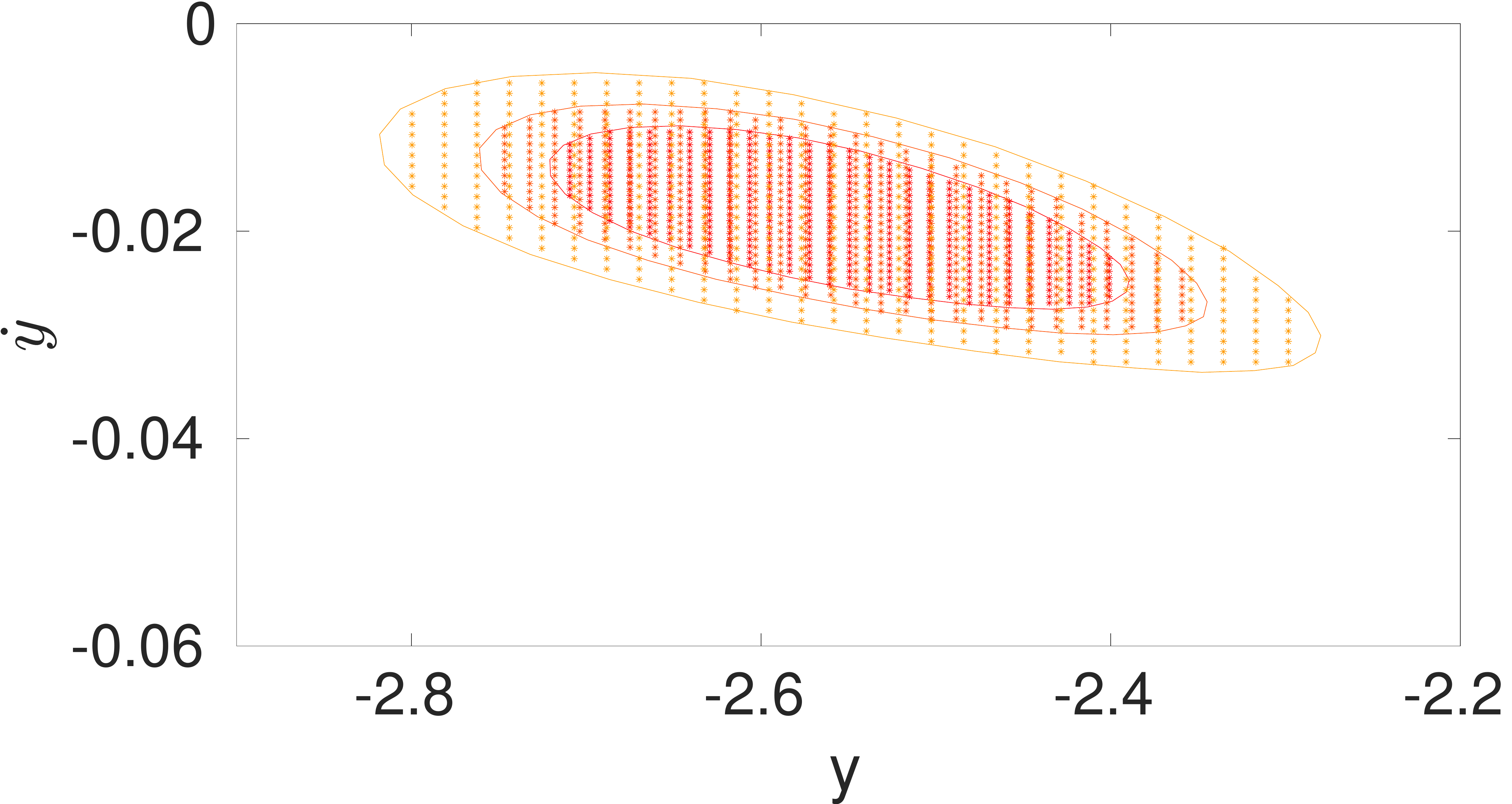}
    \includegraphics[width=0.245\textwidth]{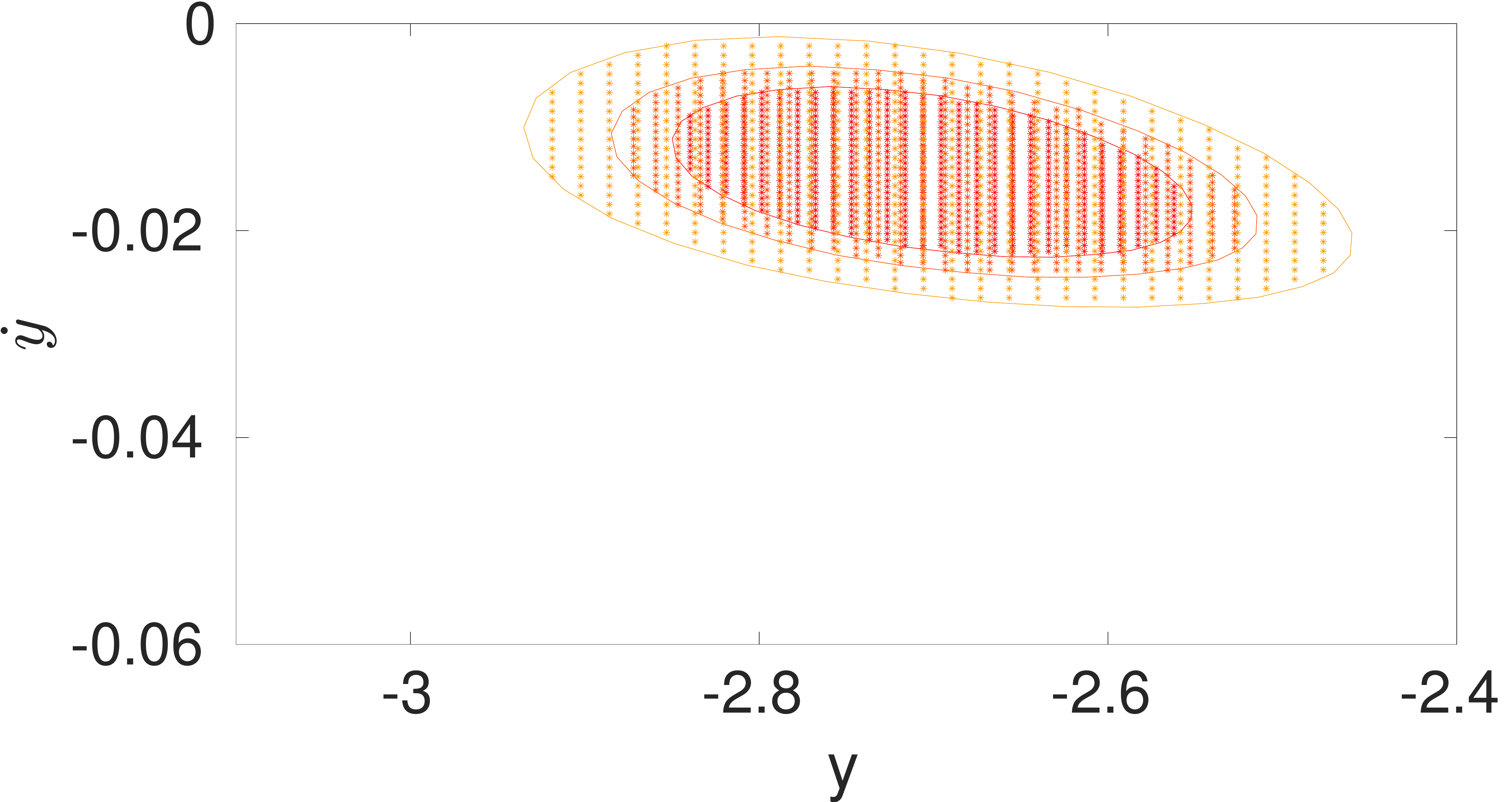}\includegraphics[width=0.245\textwidth]{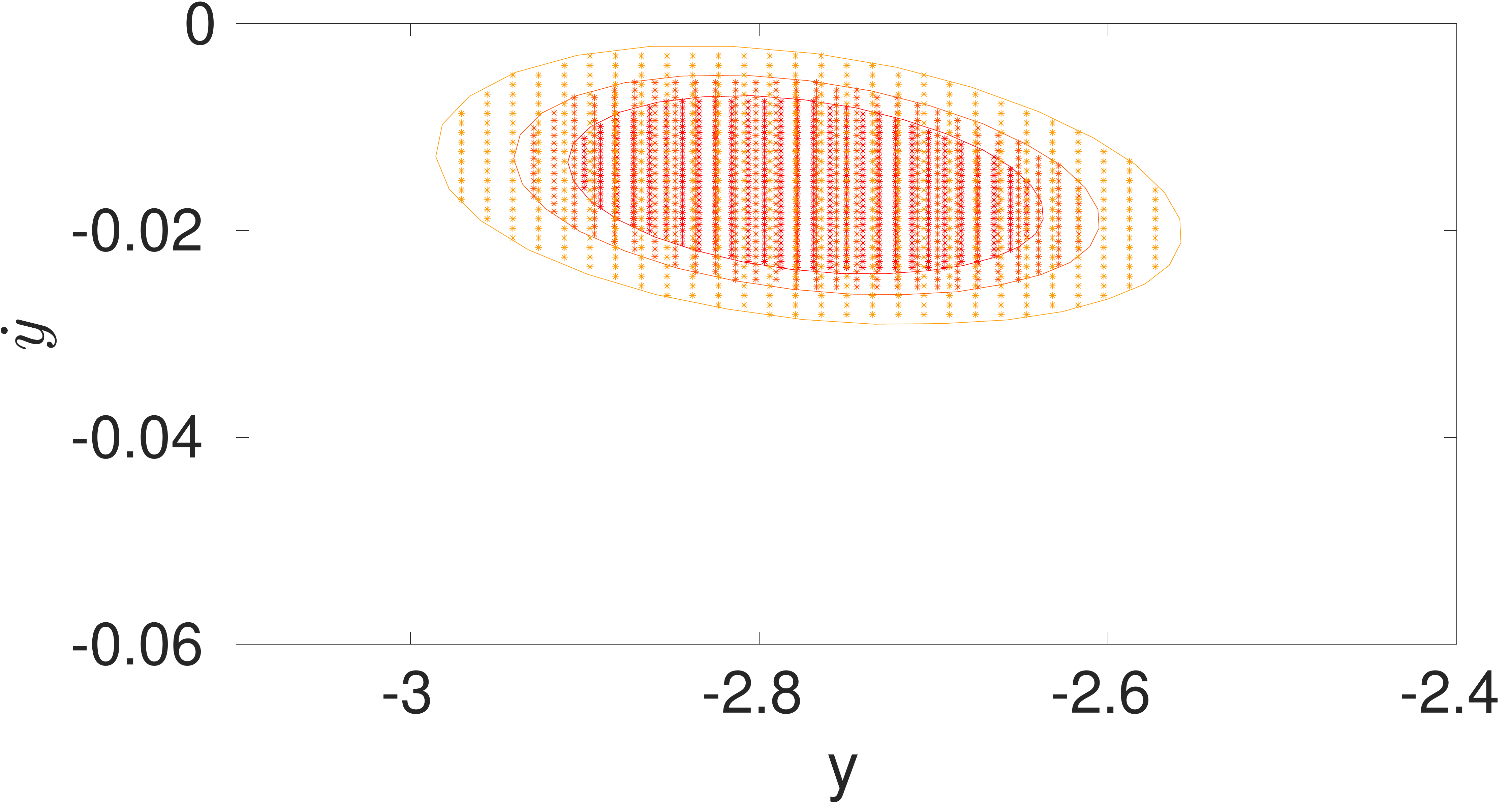}

  \caption{$(y,\,\dot{y})$ projection of the intersection of the plane $S$ with the stable manifolds associated to three orbits with different Jacobi constant of the Lyapunov family around $L_2$ (top) and $L_1$ (bottom). The intersection of each manifold is in a color, from red to yellow, according to the energy of the manifold. Initial conditions distributed inside each curve are marked with a cross with the same color as the curve. From left to right: Bulge centered at $(0,0,0)$, $(0.5,0,0)$, $(1,0,0)$ and $(1.5,0,0)$. Note that the axis limits are different but their scale and range length are constant in each row.}
  \label{fig:corteLi}
\end{figure*}

\begin{figure*}
  \centering
    \includegraphics[width=0.245\textwidth]{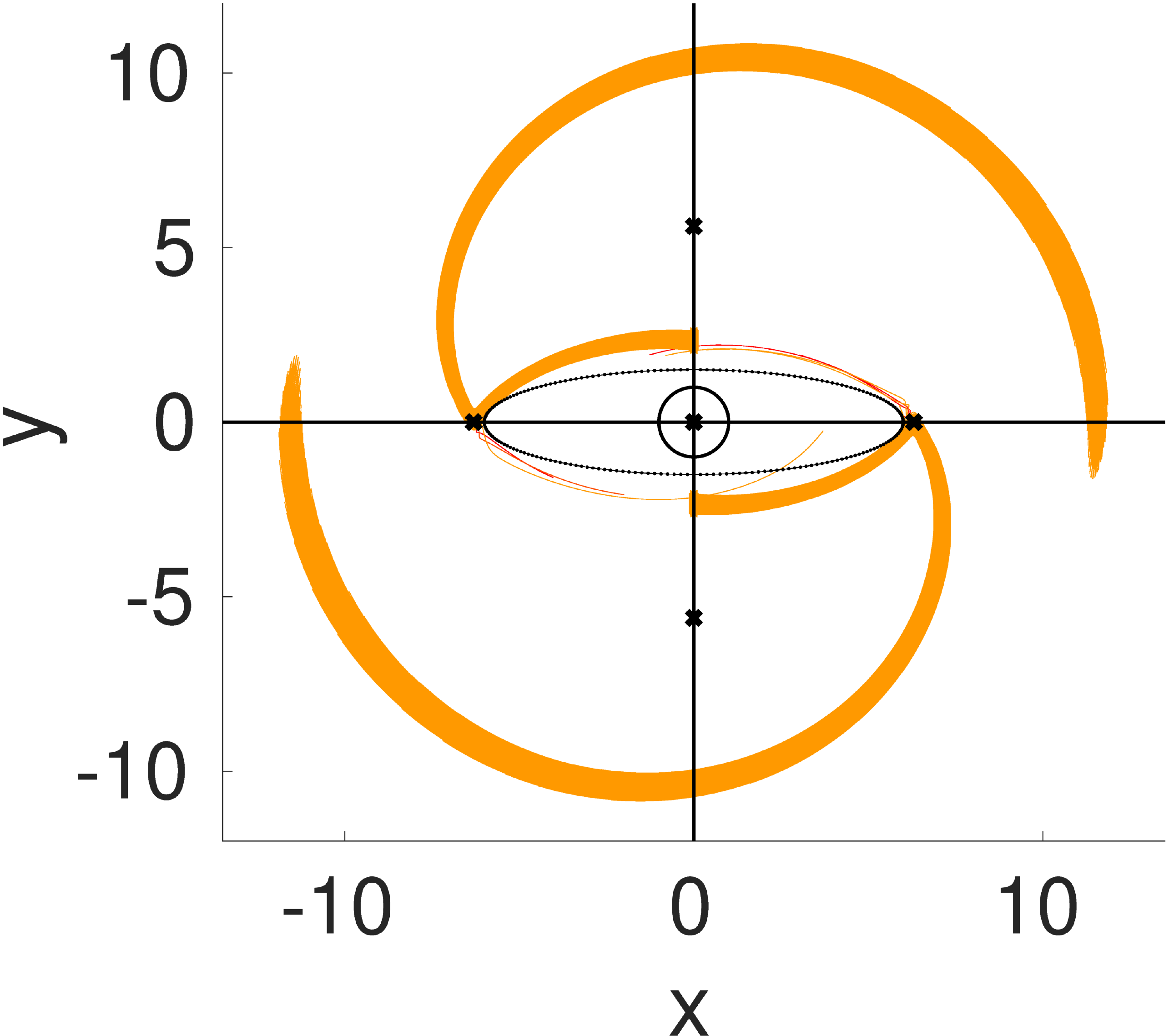} 
    \includegraphics[width=0.245\textwidth]{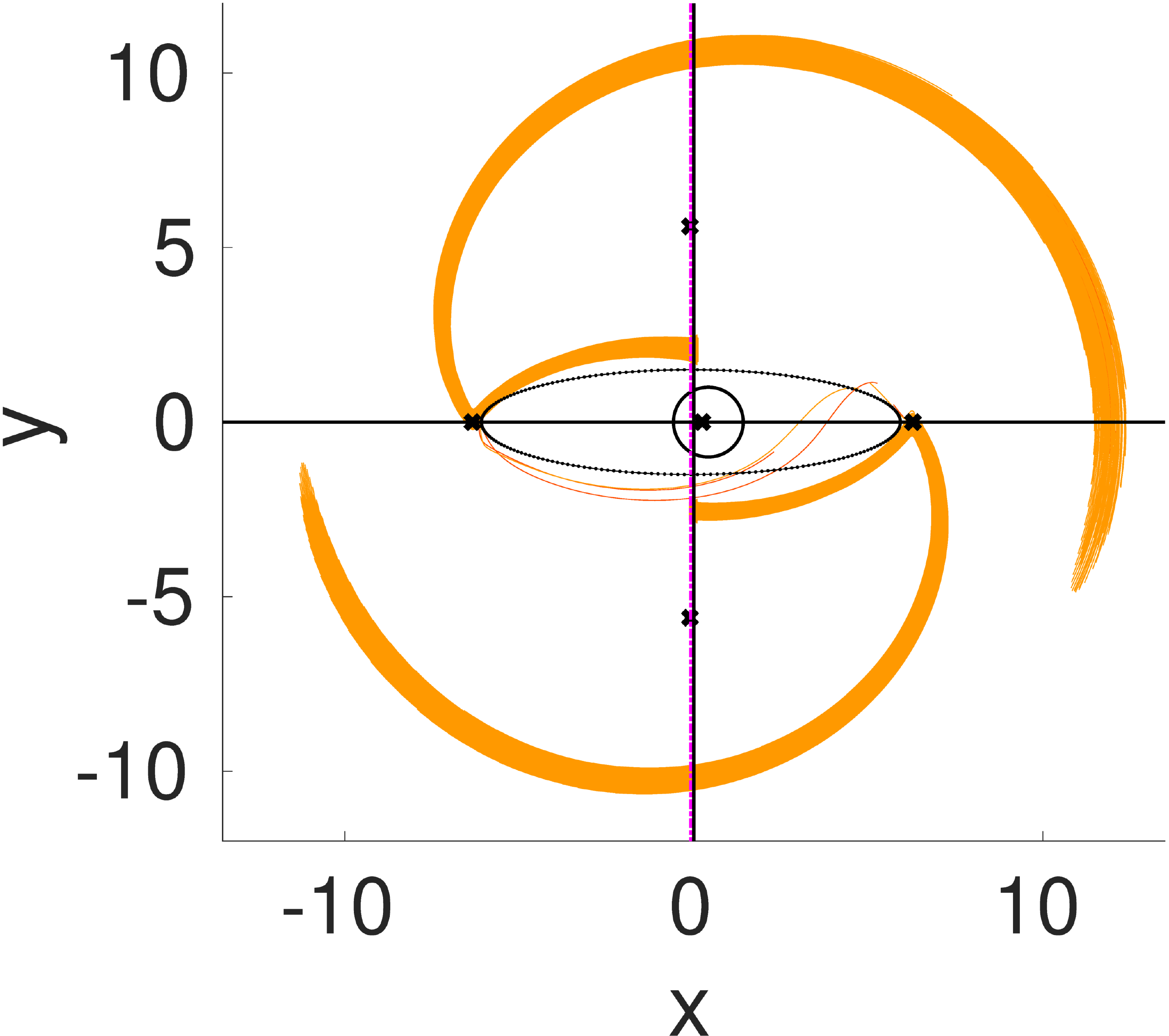}
    \includegraphics[width=0.245\textwidth]{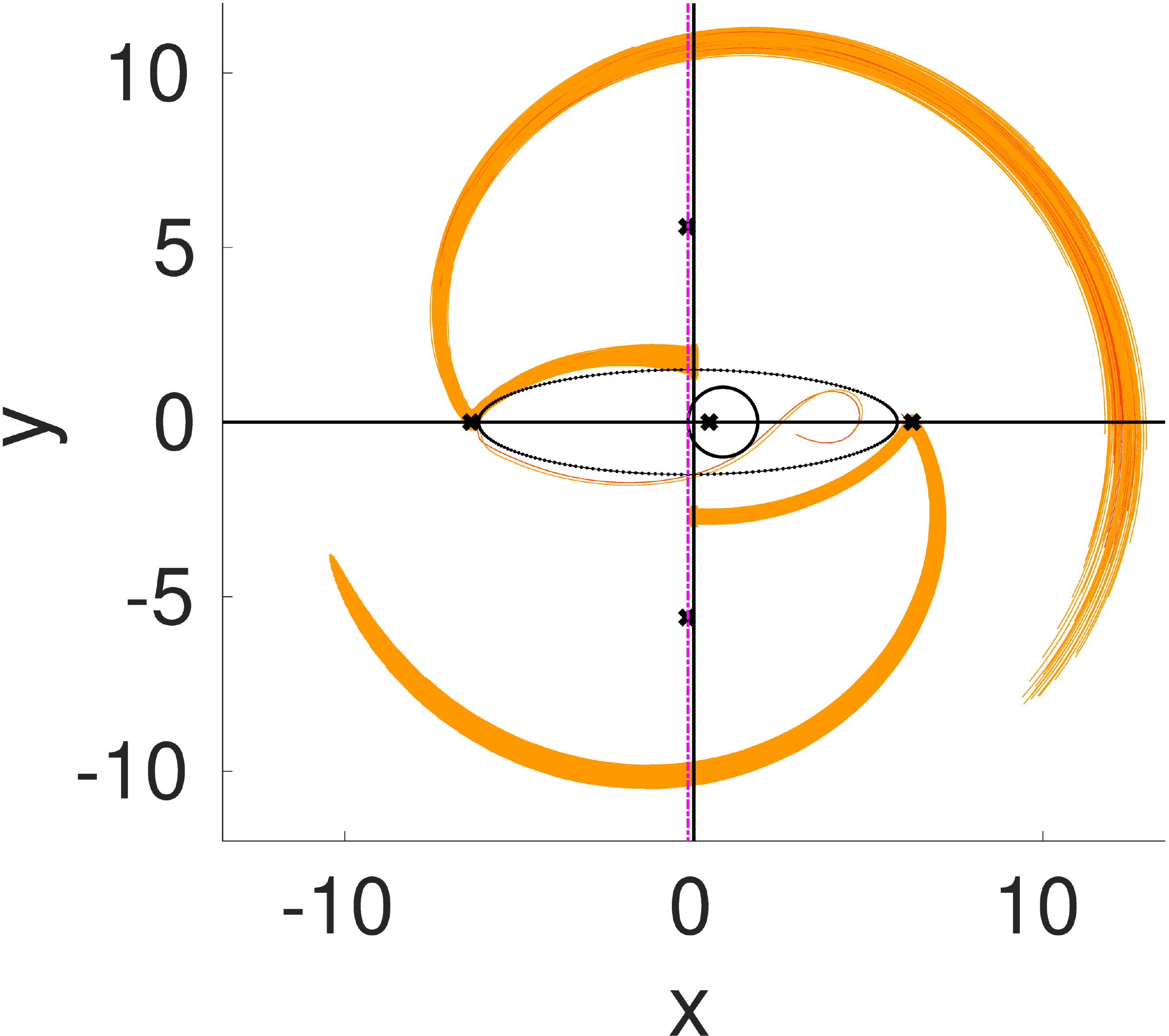} 
    \includegraphics[width=0.245\textwidth]{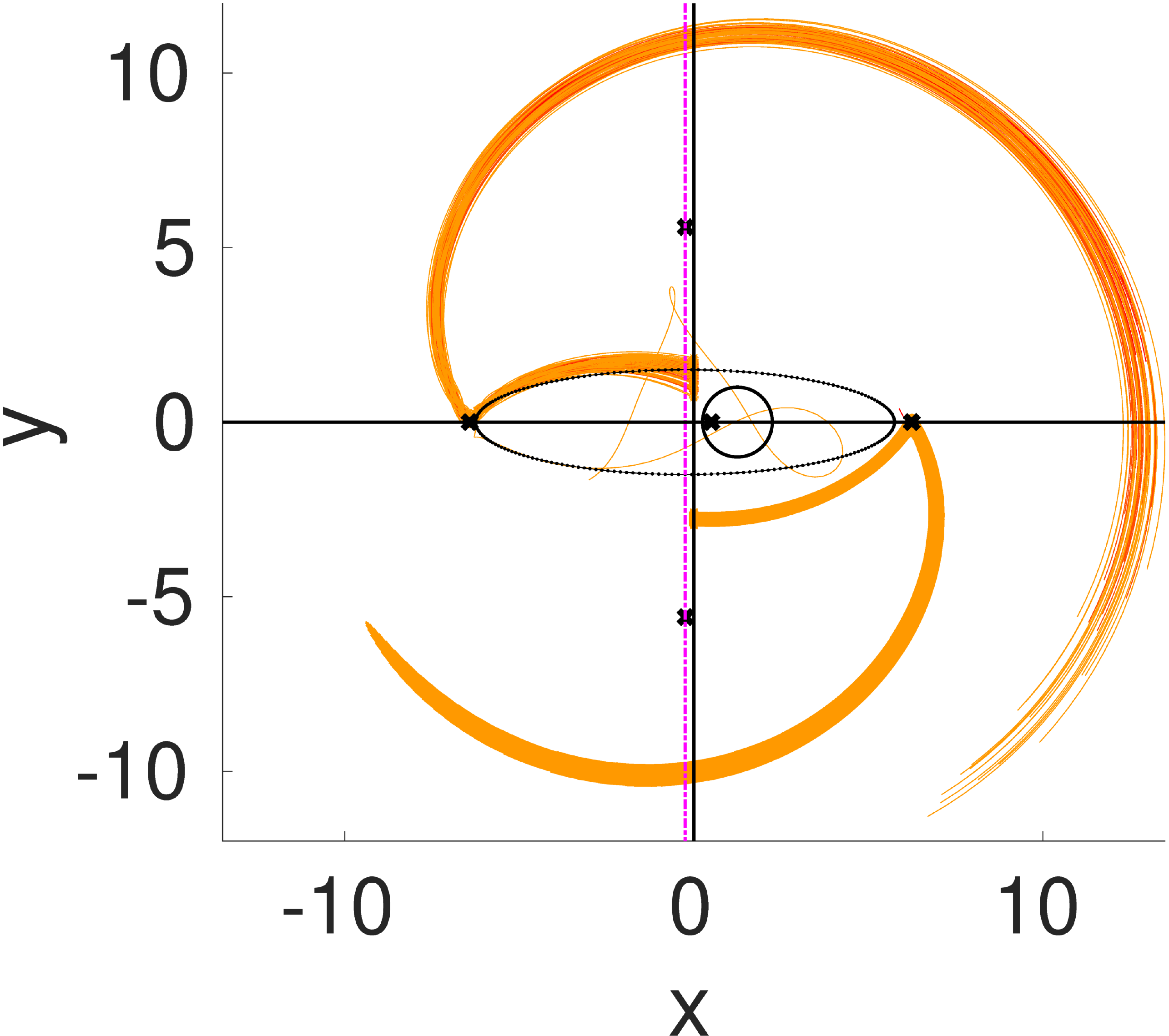}
    
  \caption{Orbits resulting from the integration of the initial conditions of Fig.~\ref{fig:corteLi}. Equilibrium points marked in red. Bar and bulge outlined by dotted black curves. The reference system is marked with a solid black line and the center of the bar with a dotted magenta line. From left to right: Bulge centered at $(0,0,0)$, $(0.5,0,0)$, $(1,0,0)$ and at $(1.5,0,0)$.}
  \label{fig:orbits}
\end{figure*}

This procedure enables us to quantify the amount of matter transferred inside each invariant manifold associated to any energy level and, consequently, from each spiral arm of the galaxy. Fig.~\ref{fig:corteLi} represents the $(y,\,\dot{y})$ projection of the intersection with the hyperplane $S$ of the invariant manifolds arising from three orbits with different Jacobi constants in the Lyapunov family around $L_2$ (top), and the same corresponding orbits for $L_1$ (bottom). The three closed curves are displayed with different colors, from red to yellow, according to the increasing Jacobi constant. The initial conditions located inside each of the closed curves are marked with crosses of the same color as the curve. From left to right, the figure displays the $(y,\,\dot{y})$ projection for the models with bulge center ranging from $(0,0,0)$ to $(1.5,0,0)$. The main feature to notice in this figure is the narrowing and stretching of the curves as the bulge moves away from the $L_2$ equilibrium point. This constriction marks the difference between initial conditions for the escape orbits related with $L_2$ and those related to $L_1$. The initial conditions emanating from near $L_2$ are more extended along the $y$-axis, resulting in a dispersion in space of the orbits enclosed by the manifold. As we integrate forward in time these initial conditions, we obtain fewer escape orbits, and dispersed in a wider region, in comparison to those emanating near $L_1$. So, the spiral arm defined by these orbits becomes more spread out and less bright. The result of these integrations is exhibited in Fig.~\ref{fig:orbits} where it can be appreciated how the orbits inside the invariant manifolds develop the arms. The orbits corresponding to the $L_1$ point are more concentrated as the density of the bar increases in the region close to $L_1$.

\section{Discussion and conclusions}
\label{sec:disc}

The goal of this work is to analyze the relation between the asymmetric arms observed in certain barred galaxies and the mass distribution in the central part of the galaxy. We propose and show that there is a strong correlation between their asymmetries. The models used to study this fact consist of a superposition of a bar and an off-centered bulge. The displacements of the bulge along the main axis of the bar introduce an increasing asymmetry in the central part of the model. When the center of mass of the galaxy is displaced along the major axis of the bar, the zero velocity curves of the effective potential become asymmetric: the one at the opposite side of the dispacement becomes more open (see Fig.~\ref{fig:cvz}). We show that this asymmetry between the zero velocity curves in their opening is carried to the orbits trapped by invariant manifolds around the Lagrangian equilibrium points $L_1$ and $L_2$. This orbits become asymmetric as well, with the arm at the opposite side of the displacement wider and more spread out. We quantify this effect by computing the fraction of orbits trapped by the manifolds. The procedure consists in intersecting the manifold with a hyperplane and, inside the resulting closed curve in the $(y,\,\dot{y})$ projection, obtaining the set of points with the same energy of the manifold as initial conditions that characterize the escaping orbits. 

Indeed, the above closed curve turns out to be the most relevant feature in order to predict the asymmetry of the arms. When the bulge is displaced along the main axis of the bar, causing an asymmetry in the density distribution of the model, the closed curves around the equilibrium point at the end of the less dense side of the bar, narrow. This leads to a smaller measure of states in phase space that become initial conditions for escaping orbits. Moreover, these closed curves are more stretched, making the distribution of the points inside them more spread out in space. Both aspects are responsible for the resulting ragged and dispersed spiral arms in the less dense end of the bar, while the arm associated with the denser end becomes brighter and well defined. 

The dynamics of off-centered bars have been studied using analytical models \citep{Colin1989}, showing how the Lagrangian points vary as a function of the displacement with respect to the center of mass of the galaxy. \citet{Lokas2021} digs into the barred galaxies in the IllustrisTNG simulations to check how common asymmetric and off-centered bars are and study its possible origin, concluding that asymmetric bars are persistent in time and this asymmetric may be due to the interaction with a companion galaxy or due to the disc itself being asymmetric. In either case, no further development has been proposed to link the asymmetry in the bar with the one-armed dominant spiral structure.

As shown in Fig.~\ref{fig:Barred}, this is a quite common phenomenon in barred galaxies. In particular, the dynamics of NGC~1300 have been studied in detail by \citep{Patsis2010}, including an orbital analysis. The isocontours of a smooth K-band image (see their Fig.~1) show a clear asymmetric bar distribution leading to an asymmetry in the spiral arms, which is reproduced by the different orbital models. Other examples of one armed dominated galaxies with an asymmetric bar may be: NGC~4027 \citep{Phookun}, the density contours show an asymmetric and off-centered bar leading to a mass distribution dominated by an $m=1$ mode, though a weak counter-part is also clear; the Large Magellanic Cloud \citep{deVaucouleurs1972}, showing a rotational asymmetry, which is confirmed by more recent studies \citep[e.g.][and references therein]{JimenezArranz22, Niederhofer22}. 

To sum up, we show that asymmetric arms are a common feature in barred galaxies and that there is a clear correlation between arm asymmetry and the displacement of the center of mass caused by an asymmetric central density distribution. Escaping orbits trapped in the invariant manifolds of an asymmetric bar distribution are asymmetric and the limiting case would be to have only one armed barred spiral.


\section*{Acknowledgements}

P.S.M. thanks the Spanish Ministry of Economy grants PID2020-117066GB-I00 and PID2021-123968NB-I00.
J.J.M. thanks MINECO-FEDER for the grant PID2021-123968NB-I00 and the Catalan government grant 2017SGR-1049.
M.R.G. thanks the Spanish Ministry of Science grant MICIU/FEDER RTI2018-095076-B-C21, and the Institute of Cosmos Sciences University of Barcelona (ICCUB, Unidad de Excelencia "Mar\'{\i}a de Maeztu") grant CEX2019-000918-M.

Funding for the Sloan Digital Sky Survey IV has been provided by the Alfred P. Sloan Foundation, the U.S. Department of Energy Office of Science, and the Participating Institutions. SDSS (www.sdss.org) acknowledges support and resources from the Center for High-Performance Computing at the University of Utah.

\section*{Data Availability}

The data underlying this article are available in the article.



\bsp	
\label{lastpage}
\end{document}